\documentclass{aa}  

\usepackage[]{graphicx} 
\usepackage{lscape}
\usepackage{txfonts}
\usepackage{array}
\newcolumntype{M}{>{\scriptsize\centering\arraybackslash}m{1.6cm}}
\newcolumntype{N}{>{\scriptsize\centering\arraybackslash}m{0.6cm}}
\begin{document}

   \title{Luminous red nova AT~2019zhd, a new merger in M~31}

 \subtitle{Forbidden hugs in pandemic times - I}

   \author{A. Pastorello\inst{1}
          \and
            M. Fraser\inst{2}
          \and
            G. Valerin\inst{1,3}
          \and
            A. Reguitti\inst{4,5,1}
          \and
            K. Itagaki\inst{6}
          \and            
            P. Ochner\inst{3}
          \and
            S.~C. Williams\inst{7,8}
          \and
             D.~Jones\inst{9,10}
          \and
            J.~Munday\inst{9,11}  
          \and
       S.~J.~Smartt\inst{12}
           \and
          K.~W.~Smith\inst{12}
         \and
          S.~Srivastav\inst{12}
          \and
          N. Elias-Rosa\inst{1,13}
           \and
          E. Kankare\inst{8}
         \and
          E.~Karamehmetoglu\inst{14}    
           \and
          P.~Lundqvist\inst{15}
          \and
            P.~A. Mazzali\inst{16,17}
           \and
          U. Munari\inst{1}
           \and
          M.~D. Stritzinger\inst{14}
          \and
          L.~Tomasella\inst{1}
           \and
          J.~P. Anderson\inst{18} 
            \and 
          K.~C.~Chambers\inst{19}
           \and
          A.~Rest\inst{20,21} 
 }

\institute{
    INAF - Osservatorio Astronomico di Padova, Vicolo dell'Osservatorio 5, I-35122 Padova, Italy  \email{andrea.pastorello@inaf.it}
         \and
    School of Physics, O’Brien Centre for Science North, University College Dublin, Belfield, Dublin 4, Ireland
         \and
    Universit\'a degli Studi di Padova, Dipartimento di Fisica e Astronomia, Vicolo dell’Osservatorio 2, 35122 Padova, Italy 
         \and
  Departamento de Ciencias Fisicas, Universidad Andres Bello, Fernandez Concha 700, Las Condes, Santiago, Chile
         \and
    Millennium Institute of Astrophysics (MAS), Nuncio Monsenor S\'otero Sanz 100, Providencia, Santiago, Chile
           \and
    Itagaki Astronomical Observatory, Yamagata 990-2492, Japan
        \and
   Finnish Centre for Astronomy with ESO (FINCA), Quantum, Vesilinnantie 5, University of Turku, FI-20014 Turku, Finland
        \and
   Department of Physics and Astronomy, University of Turku, FI-20014 Turku, Finland
          \and
  Instituto de Astrof\'isica de Canarias, E-38205 La Laguna, Tenerife, Spain
         \and
  Departamento de Astrof\'isica, Universidad de La Laguna, E-38206 La Laguna, Tenerife, Spain
         \and
           Astrophysics Research Group, Faculty of Engineering and Physical Sciences, University of Surrey, Guildford, Surrey, GU2 7XH, United Kingdom
          \and
             Astrophysics Research Centre, School of Mathematics and Physics, Queen’s University Belfast, BT7 1NN, UK
          \and
             Institute of Space Sciences (ICE, CSIC), Campus UAB, Carrer de Can Magrans s/n, E-08193 Barcelona, Spain
           \and
             Department of Physics and Astronomy, Aarhus University, Ny Munkegade 120, 8000 Aarhus C, Denmark
          \and
             The Oskar Klein Centre, Department of Astronomy, Stockholm University, AlbaNova, SE-10691 Stockholm, Sweden
          \and
             Astrophysics Research Institute, Liverpool John Moores University, ic2, 146 Brownlow Hill, Liverpool L3 5RF, UK
          \and
             Max-Planck Institut f\"ur Astrophysik, Karl-Schwarzschild-Str. 1, D-85741 Garching, Germany
         \and
             European Southern Observatory, Alonso de Córdova 3107, Casilla 19, Santiago, Chile
          \and
           Institute for Astronomy, University of Hawaii, 2680 Woodlawn Drive, Honolulu, HI 96822, USA
           \and
          Space Telescope Science Institute, 3700 San Martin Drive, Baltimore, MD 21218, USA
          \and
          Department of Physics and Astronomy, Johns Hopkins University, Baltimore, MD 21218, USA
 }


  \abstract
   {We present the follow-up campaign of the luminous red nova (LRN) AT~2019zhd, the third event of this class observed in M~31. 
The object was followed by several sky surveys for about five months before the outburst, during which it showed a slow luminosity rise. In this phase, the absolute magnitude ranged from   $M_r=-2.8\pm0.2$ mag to  $M_r=-5.6\pm0.1$ mag. Then, over a four to five day period,
AT~2019zhd experienced a major brightening, reaching a peak of $M_r=-9.61\pm0.08$ mag and an optical luminosity of $1.4 \times 10^{39}$~erg~s$^{-1}$. After a fast decline, the light curve settled onto a short-duration plateau in the red bands. Although less pronounced, this feature is reminiscent of the second red maximum observed in other LRNe. This phase was followed by a rapid linear decline in all bands.
At maximum, the spectra show a blue continuum with prominent Balmer emission lines. The post-maximum spectra show a much redder continuum, resembling that of an intermediate-type star. In this phase, H$\alpha$ becomes very weak, H$\beta$ is no longer detectable, and 
 a forest of narrow absorption metal lines now dominate the spectrum. The latest spectra, obtained during the post-plateau decline, show a very red continuum ($T_{eff} \approx$~3000~K) with broad molecular bands of TiO, similar to those of M-type stars. 
The long-lasting, slow photometric rise observed before the peak resembles that of LRN V1309~Sco, which was interpreted as the signature of the common-envelope ejection. The subsequent outburst is likely due to the gas outflow following a stellar merging event. The inspection of archival HST images taken 22 years before the LRN discovery reveals a faint red source ($M_{F555W}=0.21\pm0.14$ mag, with $F555W-F814W = 2.96\pm0.12$ mag) at the position of AT~2019zhd, which is the most likely quiescent precursor. The source is consistent with expectations for a binary system including a predominant M5-type star.
}


   \maketitle

\section{Introduction}

The luminous red nova (LRN) designation covers a heterogeneous class of gap transients \citep{pasto19}. In contrast with other families of gap 
transients, LRNe span an enormous luminosity range. Intrinsically faint transients ($M_V \approx -4$ to $-6.5$ mag) such as OGLE 2002-BLG-360 
\citep{tyl13} and V1309~Sco \citep{mas10,tyl11} are usually observed in our Galaxy, while much brighter events ($M_V \lesssim -12$ mag) 
such as NGC4490-2011OT1 \citep{smi16,pasto19a}, AT~2015dl in M~101 \citep{bla17}, AT~2017jfs \citep{pasto19b}, and AT~2018hso \citep{cai19} are 
occasionally discovered in other galaxies. This suggests that faint LRNe are common and can be observed with a rate of
once per decade in a galaxy such as the Milky Way, while luminous events are intrinsically rare \citep{koc14,how20}.

The detailed photometric and spectroscopic follow-up of V1309~Sco revealed the physical mechanisms powering at least most of LRN outbursts. 
For a few years before the outburst, V1309~Sco showed a 
short-period photometric modulation with a pseudo-period of about 1.4~d due to the inspiraling motion of the secondary
star in a binary system. Later, the photometric period started to decline. The photometric modulation
disappeared as soon as the light curve reached a minimum followed by a slow luminosity rise of $\sim$4 mag, lasting about six months. 
This was interpreted as the signature of the common-envelope (CE) ejection which obscured the binary system. The final LRN outburst
was likely due to the gas outflow following the stellar coalescence. Recent theoretical studies \citep[e.g.][]{pej16,pej17,mac17,mac18} 
support this scenario. 

Starting from an alternative characterization of the sub-families of gap transients, \citet{sok20b} recently proposed a 
jet-powered toy model to explain their observed properties of different types of objects.
Regardless of whether the physical scenario is a merging event or simple mass transfer that leaves
the binary system intact, the accretion of mass through a disk may launch polar jets \citep{kas16,sok16}.
The efficient conversion of kinetic energy into radiation produced by the jet colliding with pre-existing circum-stellar material 
may account for the light curves of LRNe and other gap transients \citep[see, also,][]{sok20a}.

The large range of observed peak luminosity of LRNe is related to the different masses involved.
\citet{koc14} propose that the most luminous outbursts are produced by more massive mergers. This finding is supported by
\citet{pasto19a}, who found a possible correlation between the luminosity of individual light curve features and the luminosity 
of the quiescent progenitor system.

   \begin{figure}
   \centering
   \includegraphics[width=8.8cm,angle=0]{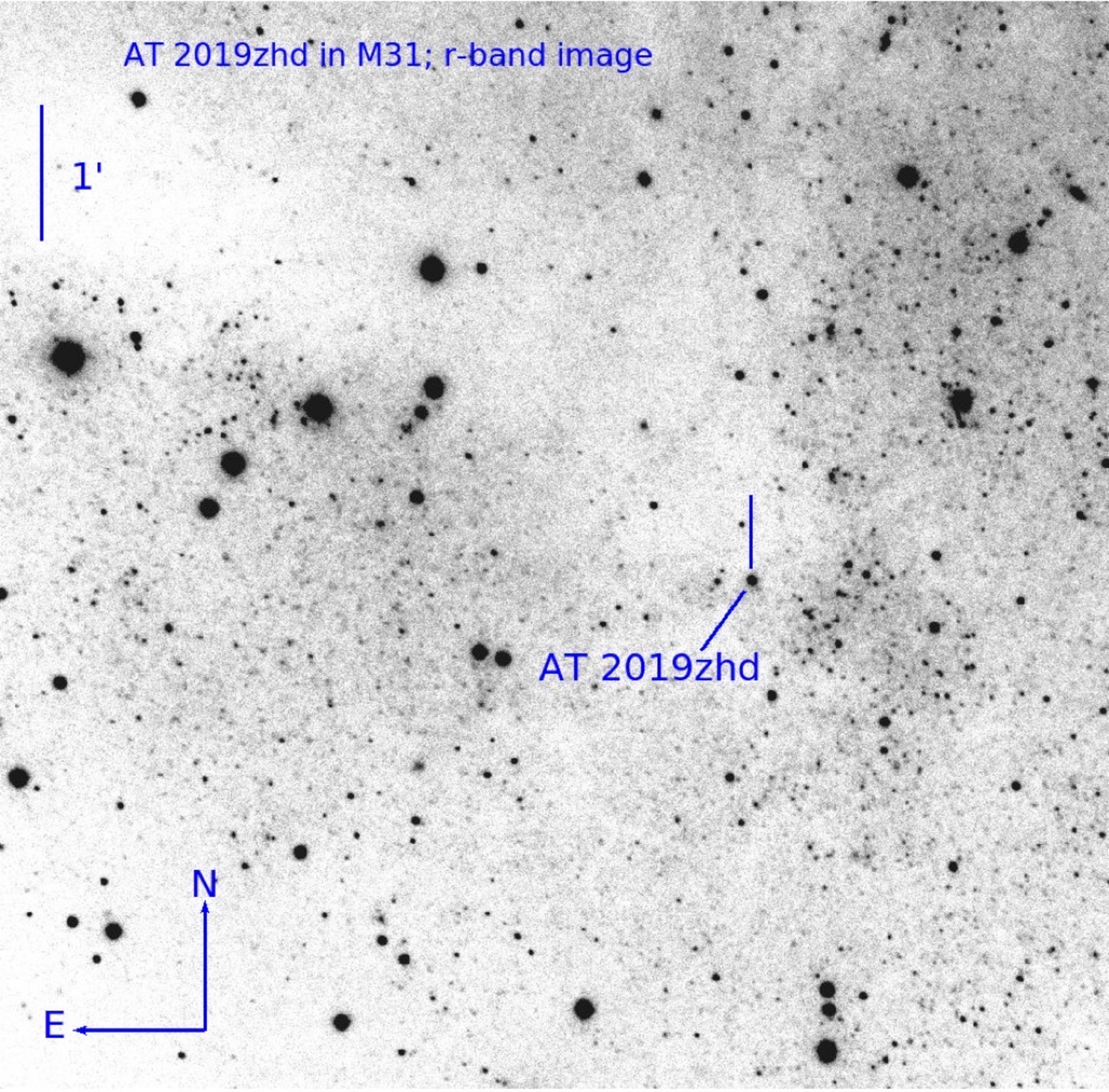}
   \caption{AT~2019zhd in M~31: $r$-band image of the LRN field obtained on 2020 February 14 with the 1.82m Asiago Copernico telescope, equipped with AFOSC.}
              \label{fig1}
    \end{figure}

In this paper, we present the results of our follow-up campaign of LRN AT~2019zhd, which was discovered in M~31. This is the third LRN 
that has been discovered so far in M~31, after M31-RV \citep{bos04} and M31-LRN2015 \citep{kur15,wil15,lip17,bla20}. Interestingly,
all of them have rather similar observational properties. The follow-up campaign of AT~2019zhd has been
carried out during the pandemic period in part, and has been affected by a limited access to the observational facilities. 
Despite these restrictions, the past few months have been prolific in terms of LRN discoveries as AT~2020hat and AT~2020kog 
\citep{pasto20} were announced soon after AT~2019zhd.

The format of the paper is as follows. In Sect. \ref{host}, we report information on the discovery of AT~2019zhd, along with 
its distance and interstellar reddening; in Sect. \ref{photometry}, we discuss the photometric properties during the different phases
of the LRN evolution, while in Sect. \ref{spectroscopy} we describe the spectroscopic evolution; in Sect. \ref{radius} we analyse the bolometric
light curve and the evolution of the photospheric temperature and radius, compared with the other LRNe in M~31; in Sect. \ref{progenitor} we characterise the nature of
the progenitor system, while a discussion and a short summary follow in Sect. \ref{discussion}.

\section{The discovery of AT~2019zhd} \label{host}

AT~2019zhd\footnote{Alternative survey designations are ZTF19adakuot and ATLAS19berq.} 
was discovered by the Zwicky Transient Factory (ZTF) on 2019 December 14.10 UT \citep{ho19} at an apparent magnitude
$r=20.35 \pm 0.20$ mag. The transient is observed at  RA = $00^{h}40^{m}37\fs905$ and
 Dec = $+40\degr34\arcmin52\farcs85$ (equinox J2000.0), which is in the south-west outskirts of the M~31 disk.
The object was then spectroscopically classified as a possible luminous blue variable (LBV) outburst or a LRN by \citet{kaw20}. 
The association of the transient with the galaxy is confirmed through its spectra, whose emission lines are slightly shifted towards
bluer wavelengths, consistent with the negative recessional velocity of M~31\footnote{The redshift of AT~2019zhd measured from our early
spectra (see Sect. \ref{spectroscopy}) is consistent with that of the nearby emission line source 2035 ($z=-0.001764$) discussed by \citet{mer06}.}.
The region in M~31 with AT~2019zhd is shown in Fig. \ref{fig1}.

The distance of M~31 is well constrained through a number of methods. In this paper, we adopt the statistical estimate
from \citet{tul13}, based on different methods (including cepheids, the tip of the red giant branch, and the surface brightness fluctuation)
and scaled to H$_0$ = 73 km s$^{-1}$ Mpc$^{-1}$. The inferred distance is $d=0.785\pm0.009$ Mpc, which 
provides a distance modulus $\mu = 24.47 \pm 0.06$ mag. 

The Galactic line-of-sight extinction at the location of the transient is $E(B-V) = 0.055$ mag \citep{sch11}.
Our early spectra of AT~2019zhd (see Sect. \ref{spectroscopy}) do not show evident absorption features of interstellar Na~I that can be
attributed to additional host galaxy reddening. For this reason, we assume that the total reddening towards AT~2019zhd is
consistent with the Galactic value. 

\section{Photometric data} \label{photometry}

   \begin{figure*}
   \centering
   \includegraphics[width=13.7cm,angle=270]{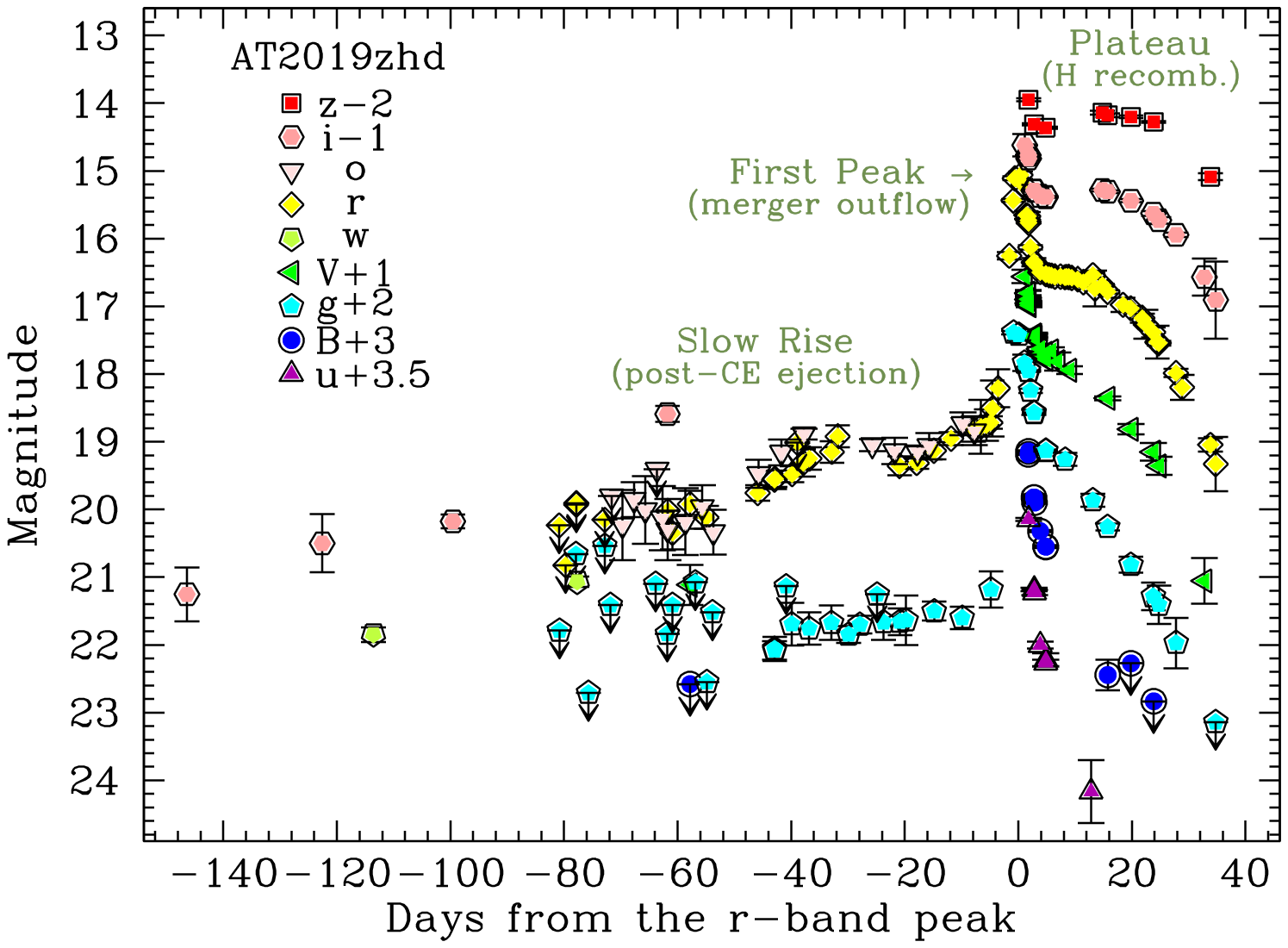}
   \caption{Long-term light curve of AT~2019zhd, which includes both pre-outburst photometry and the LRN outburst.
The phases are computed from the $r$-band maximum (MJD = 58892.0 $\pm$ 0.5). 
Only significant detection limits are shown in the figure.}
              \label{fig2}
    \end{figure*}

Photometric data reduction was carried out using the SNOoPY pipeline \citep{cap14}. After correcting the science images 
for standard calibration images (bias and flat-fields), and after performing astrometric calibration, 
SNOoPY allows us to carry out the simultaneous PSF-fitting photometry of the target along with a number of stellar sources
in the field of the transient. The instrumental photometry was then calibrated using zero points and colour term corrections
inferred, for each night and instrumental configuration, using standard stars from \citet{smi02}.
Finally, the Sloan-band photometry was fine-tuned through a comparison of the magnitudes of a number of stars in
the field of AT~2019zhd with those of the SDSS catalogue.

A catalogue of comparison stars in the  $B$ and $V$ Johnson filters was obtained
 converting SDSS to Johnson-Bessell magnitudes using the transformation relations of \citet{chr08}. 
The final magnitudes of AT~2019zhd are reported in Tables  \ref{tabA1} and \ref{tabA2}. The comprehensive multi-band light curves of AT~2019zhd
are shown in Fig. \ref{fig2}.

\subsection{Pre-outburst phase} \label{photometry_CE}

Although the object was discovered about two months before the $r$-band light-curve maximum (see Sect. \ref{photometry_LRN}), we collected data from the major public surveys 
to track its pre-outburst evolution.
Photometric data come from ZTF\footnote{The data have been obtained through the Lasair \protect\citep{smi19} and the ALeRCE brokers \protect\citep{for20}.} \citep{bel19,gra19}, 
the Panoramic Survey telescope and Rapid Response System \citep[Pan-STARRS, hereafter PS;][]{cha16,mag16}, and the Asteroid Terrestrial-impact Last Alert System \citep[ATLAS;][]{ton18,smi20}. While the ZTF photometry was obtained with nearly standard 
Sloan-$g$ and $r$ filters, PS provided data in the $i$ band (which is close to Sloan-$i$) and the wide ($w$) filter (nearly covering the $g+r$ band). 
ATLAS provided magnitudes in the cyan (hereafter $c$, which is approximately $g+r$) and orange (hereafter $o$, nearly $r+i$) filters. While the few ATLAS-$c$ magnitudes have
been converted to Sloan-$g$ magnitudes following the prescriptions of \citet{ton18}, the PS-$w$ and the ATLAS-$o$ magnitudes have been 
left in their original photometric systems.

   \begin{figure}
   \centering
   \includegraphics[width=9.4cm,angle=0]{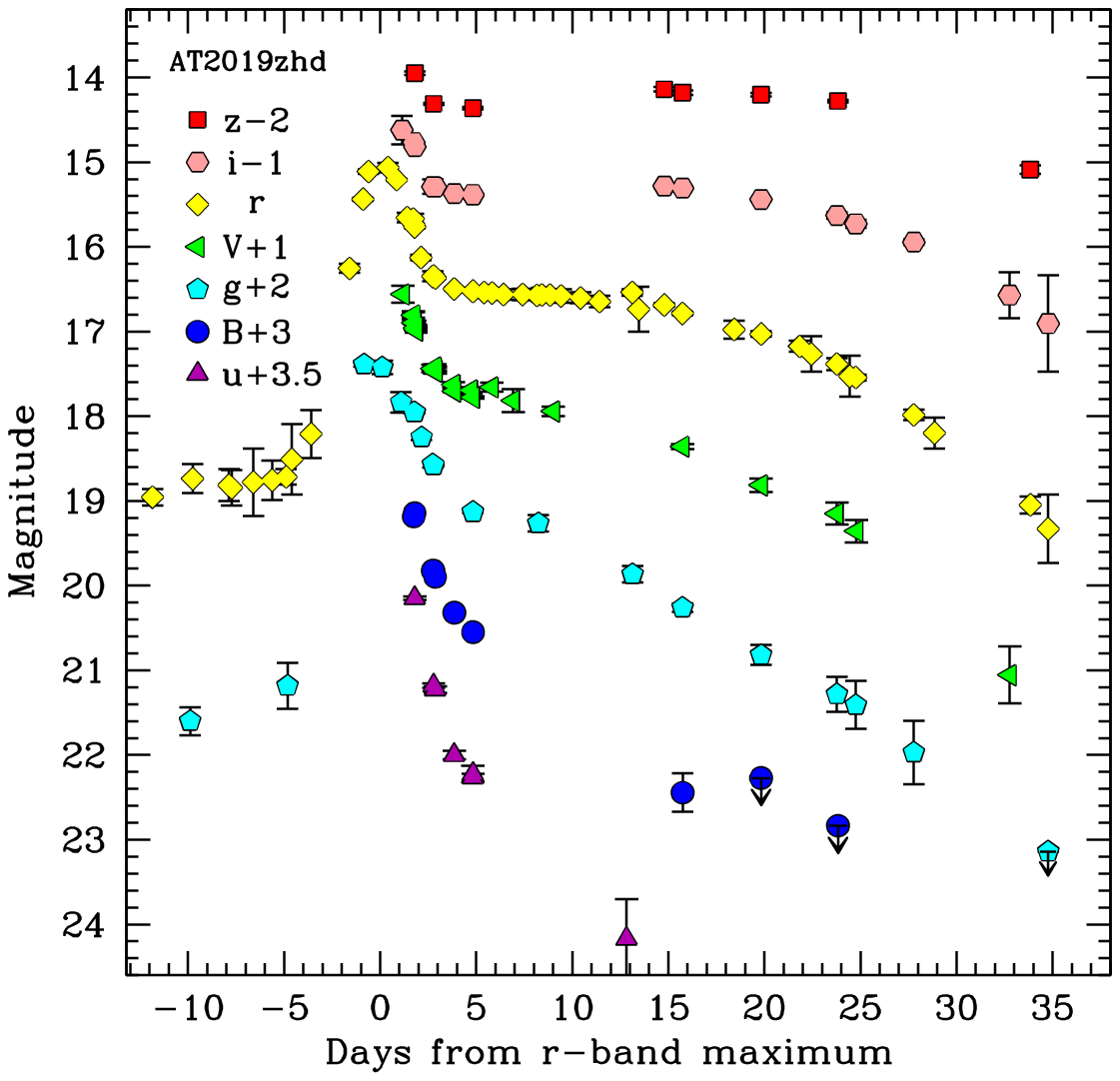}
   \caption{Multi-band light curve of the AT~2019zhd outburst.}
              \label{fig3}
    \end{figure}

 The earliest detections (from $\sim$ 150 to 100 days before maximum) are provided by the PS survey. From Fig. \ref{fig2}, we note that a very faint source (at $i=22.25\pm0.40$ mag) 
was first observed at the location of AT~2019zhd on MJD = 58745.59, hence 146 d before the peak of the outburst (see Sect. \ref{photometry_LRN}). Then, the source experienced a moderate 
magnitude rise of $1.6\pm0.3$ mag (100~d)$^{-1}$ in the $g$ band and $2.6\pm0.2$ mag (100~d)$^{-1}$ in the $r$ band (these slopes have been obtained approximating ATLAS-$o$ and PS-$w$
magnitudes to Sloan-$r$). Finally, a much more evident brightening started on MJD $\sim$ 58887; the evolution of this outburst will be detailed in Sect. \ref{photometry_LRN}.

\subsection{Multi-band light curves} \label{photometry_LRN}

The LRN outburst was well monitored by the surveys mentioned in Sect. \ref{photometry_CE} and, in the $g$ band, by the All-Sky Automated Survey for Supernovae \citep[ASAS-SN;][]{sha14,koc17}.
Supporting multi-band photometry has been obtained at the 2.0m Liverpool Telescope equipped with IO:O and the 2.56~m Nordic Optical Telescope (NOT) with ALFOSC and StanCam\footnote{The NOT data have been obtained in the framework of the Nordic optical telescope Unbiased Transient Survey 2 (NUTS2) collaboration \protect\citep{hol19}; see \url{https://nuts.sn.ie/}.}, hosted at the Roque de los Muchachos
(La Palma Canary Islands, Spain); the 1.82m Copernico Telescope with AFOSC, and the 67/92~cm Schmidt Telescope at Mt. Ekar (near Asiago, Italy). Additional unfiltered photometry (scaled to
Sloan $r$-band) has been taken using a 35~cm telescope of the Itagaki Astronomical Observatory (Yamagata, Japan).

   \begin{figure*}
   \centering
   {\includegraphics[width=9.1cm,angle=0]{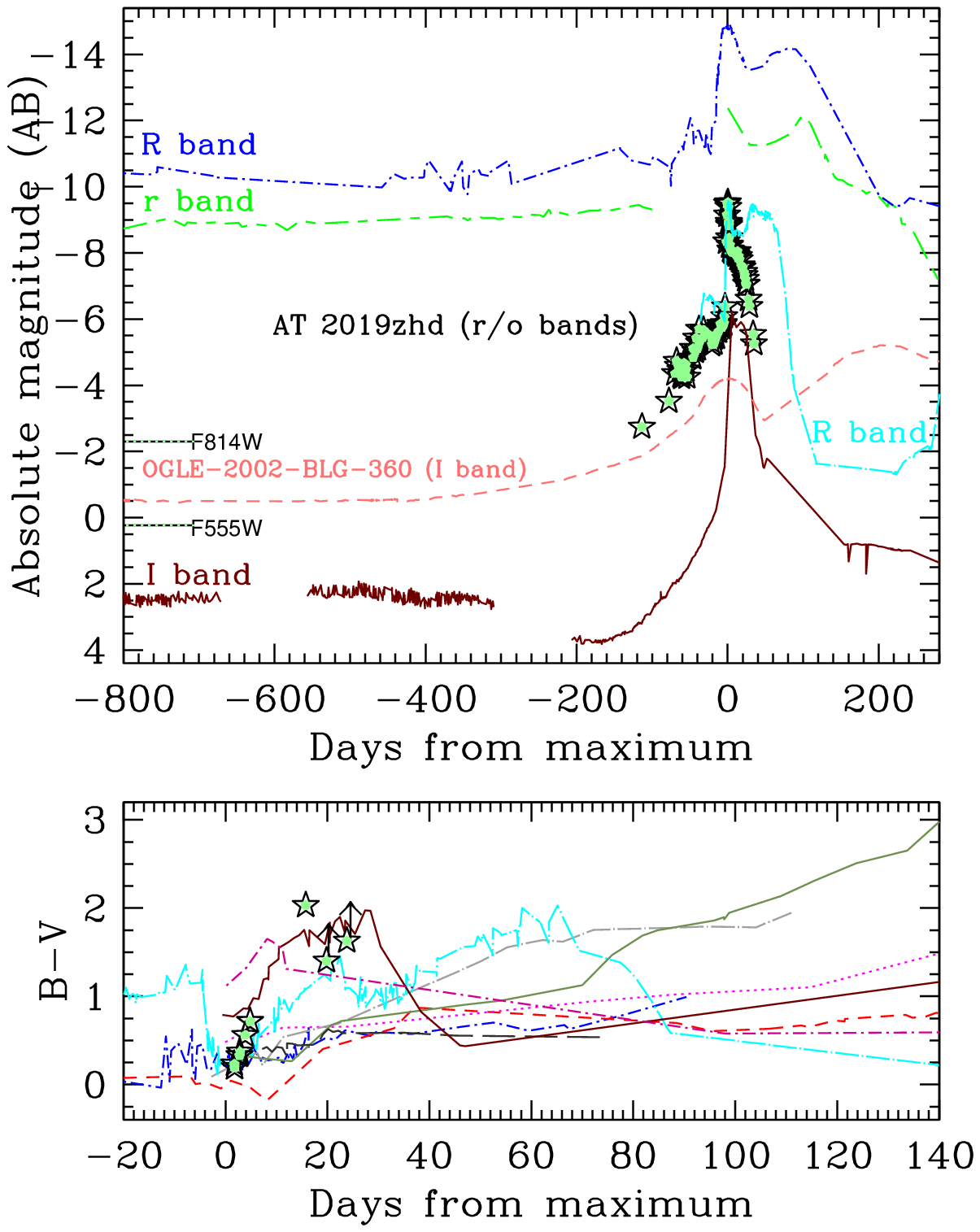}
   \includegraphics[width=9.1cm,angle=0]{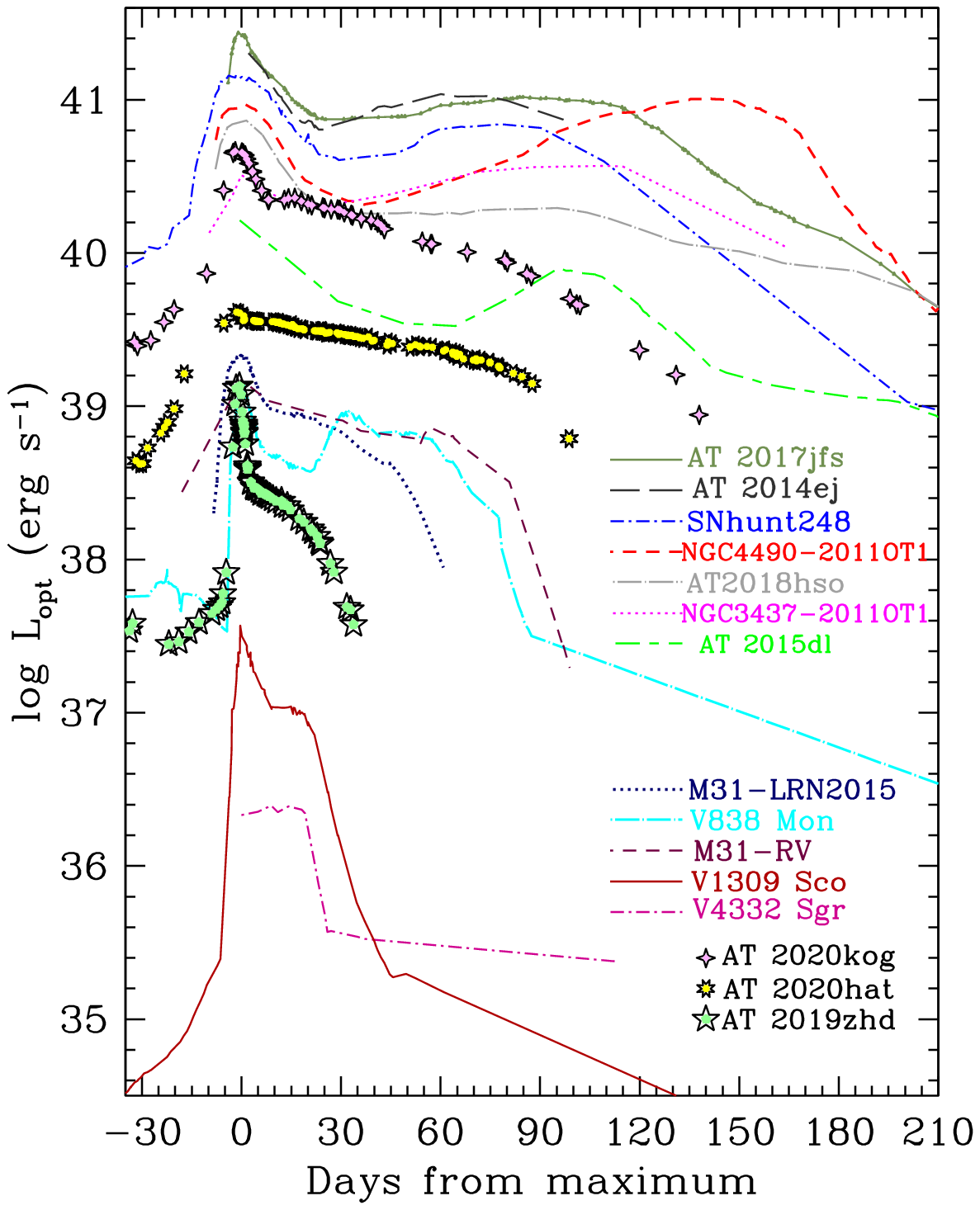}}
   \caption{{\it Top-left}: Comparison of absolute light curves for a sub-sample of LRNe, with emphasis on the pre-outburst phase. The data for the comparison objects,
observed with Johnson-Cousins filters, are reported to the ABmag system by applying a shift of +0.21 mag for the $R$-band and +0.45 mag for the $I$-band observations,
as indicated in \protect\citet{bla07}. The short dotted lines to the left indicate the absolute magnitudes of the quiescent progenitor of AT~2019zhd 
(HST-$F555W$ and $F814$ bands, in the ABmag system; see Sect. \ref{progenitor}). {\it Bottom-left}: $B-V$ colour curves of a few LRNe during the outburst. {\it Right}: pseudo-bolometric 
light curve of AT~2019zhd, along with those of a wide sample of LRNe. We used the same interstellar extinction and distance values as in \protect\citet[][see their Table 3]{pasto19a}. 
The data of AT~2018hso and AT~2014ej are from \citet[][]{cai19} and \citet{str20}, respectively, while those of AT~2020kog and AT~2020hat are from \citet{pasto20}. In all panels, 
the curves for the different objects are identified with the symbols labelled in the right panel.}
              \label{fig4}
    \end{figure*}

   \begin{table*}
      \caption[]{General information on the spectra of AT~2019zhd. The phases are from the $r$-band maximum (on MJD = 58892.0 $\pm$ 0.5).}
         \label{tab1}
     $$         \begin{array}{lllllll}
            \hline  \hline
            \noalign{\smallskip}
    $Date$ &  $MJD$ & $Phase$ & $Instrumental~configuration$ & $Exptime$ & $Resol.$ & $Range$ \\ 
         &      & $(days)$ &   & $(s)$ & $(nm)$ & $(nm)$ \\ \hline
$2020~Feb~13$ & 58892.9 & +0.9  & $LT + SPRAT$ & 300 & 1.4 & 400-800 \\
$2020~Feb~14$ & 58893.8 & +1.8  & $Copernico + AFOSC + VPH7 + VPH6$ & 1200+1200 & 1.5+1.5 & 330-930 \\ 
$2020~Feb~15$ & 58894.8 & +2.8  & $Copernico + AFOSC + VPH7$ & 2000 & 1.5 & 340-730 \\
$2020~Feb~15$ & 58894.8 & +2.8  & $Galileo + B\&C + 300tr$ & 5400 & 0.6 & 330-780 \\
$2020~Feb~17$ & 58896.9 & +4.9  & $NOT + ALFOSC + gm4$ & 1800 & 1.4 & 360-960 \\
$2020~Feb~18$ & 58897.9 & +5.9  & $Copernico + AFOSC + VPH6 + VPH7$ & 1800+1500 & 1.5+1.5 & 350-920 \\
$2020~Feb~21$ & 58900.8 & +8.8  & $GTC + OSIRIS + R2500V + R2500R$ & 360+360 & 0.21+0.25 & 443-768 \\
$2020~Feb~28$ & 58907.8 & +15.8 & $Copernico + AFOSC + VPH6$ & 3600 & 1.5 & 500-920 \\ 
$2020~Mar~05$ & 58913.9 & +21.9 & $NOT + ALFOSC + gm4$ & 2700 & 1.8 & 350-960 \\
$2020~Mar~12$ & 58920.9 & +28.9 & $NOT + ALFOSC + gm4$ & 1200 & 1.8 & 420-960 \\
\hline                    

        \noalign{\smallskip}
         \end{array}
     $$ 
\tablefoot{
    LT = 2.0~m Liverpool Telescope (La Palma, Canary Islands, Spain);
    Copernico = 1.82~m Copernico Telescope (INAF - Padova Observatory; Mt. Ekar, Asiago, Italy);
    Galileo = 1.22~m Galileo Telescope (Padova University, Asiago - Pennar, Italy);
    GTC = 10.4~m Gran Telescopio Canarias (La Palma, Canary Islands, Spain);
    NOT = 2.56~m Nordic Optical Telescope (La Palma, Canary Islands, Spain).}
   \end{table*}

The rise to the outburst peak is very fast, lasting 4-5 days in the $g$ and $r$ bands. The peak parameters were determined through a low-order polynomial fit. We obtained that the light curve
reached the peak in the $r$-band on MJD = 58892.0 $\pm$ 0.5 ($r_{max} = 15.00\pm0.01$ mag), and in the $g$-band on MJD = 58991.5 $\pm$ 0.7 ($g_{max} = 15.38\pm0.01$ mag).  Hereafter, the epoch of the $r$-band peak will be considered as
the reference time in this paper.
The maximum is followed by an initially fast decline and then by a flattening, which is marginally visible in the blue bands, while it is more evident in the redder bands. The magnitude of the plateau in the $r$ band is $r=16.55\pm0.05$ mag.
A plateau or a broad second maximum have been observed in a number of LRNe \citep[see,][and references therein]{pasto19a}. However, while the second red peak is prominent in the intrinsically
luminous events, usually it is relatively shallow in fainter LRNe \citep[see, e.g.][their figure 15]{pasto19a}. After the secondary peak (or the plateau) the light curve declines monotonically in all bands.
The overall photometric evolution of the outburst in the different bands is shown in Fig. \ref{fig3}.

\subsection{Comparison with other LRNe} \label{photometry_comp}

In Fig. \ref{fig4}, we compare the $r$-band absolute light curve, the $B-V$ colour evolution, and the pseudo-bolometric light curve of AT~2019zhd with those of a sample of LRNe \citep[see][and references therein]{pasto19a}. The top-left panel of Fig. \ref{fig4} shows the pre-outburst light curves of a few objects whose fields were covered by survey observations, highlighting the pre-outburst photometric evolution of the LRN precursors. In particular, the historical light curve of V1309~Sco
published by \citet{tyl11} is considered as a milestone for our comprehension of the LRN phenomenon. V1309~Sco is a Galactic LRN routinely observed by the OGLE project \citep{uda03} from 2001 to late 2007, and showed a modulated light curve with photometric period decreasing with time. This photometric variability was likely produced by a binary system that was rapidly losing angular momentum. In early 2008, a broad photometric minimum and a later brightening by 4 mag in about six months were observed, while signatures of a light curve modulation were no longer visible. This  was interpreted as the result of the common envelope ejection, which embedded and obscured the two stellar companions. A much more evident outburst, characterised by a brightening by further 4 mags in about one week, was later observed, and is believed to be due to gas outflow triggered by the coalescence of the two stellar cores. 

   \begin{figure*}
   \centering
   \includegraphics[scale=0.61,angle=270]{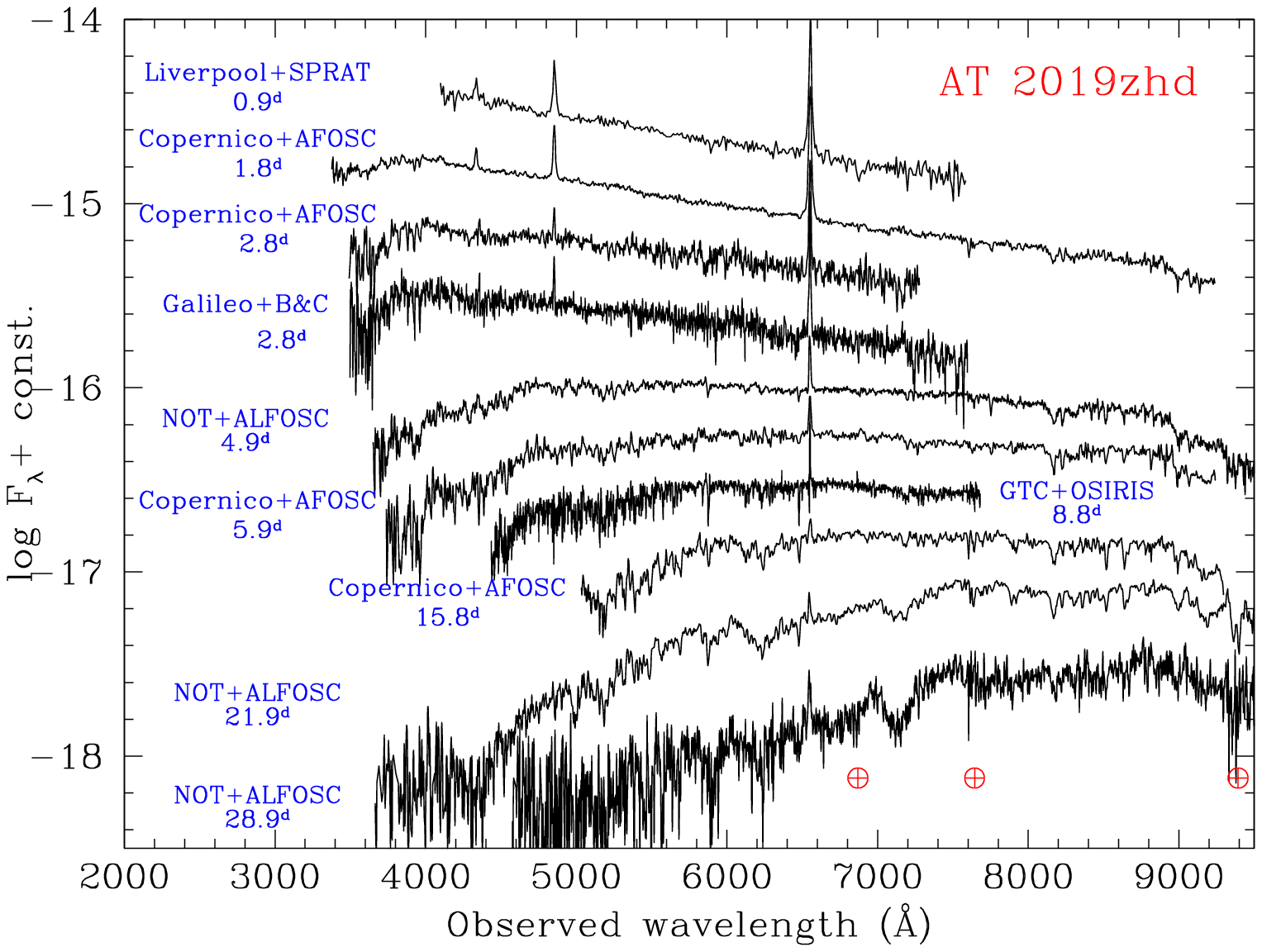}
   \caption{Spectroscopic evolution of AT~2019zhd, from the light curve peak to late phases. No reddening and redshift corrections have been applied to the spectra. 
For clarity, the spectra have been shifted in flux by an arbitrary amount. The phases  are from the Sloan-$r$ band luminosity peak.}
              \label{fig5}
    \end{figure*}

Unfortunately, V1309~Sco is the only LRN for which high signal to noise (S/N) and high cadence photometry is available. For this reason, the past history of binary variability  is not detectable for other objects. However,  
SNhunt248 \citep{kan15,mau15,mau18}, AT~2015dl \citep{bla17}, and OGLE-2002-BLG-360 \citep{tyl13} showed a long period of slowly rising luminosity before the outburst onset, with in some cases minor to moderate luminosity fluctuations  \citep[see, e.g. M31-LRN2015;][]{bla20}.

AT~2019zhd was observed for almost five months before the LRN outburst, during which its absolute magnitude increased from $M_r \approx -2.8$ to $M_r \approx -5.6$ mag. This relatively slow rise is likely due to the photometric evolution of the stellar  system after the common envelope ejection, similarly to V1309~Sco.
The light-curve bump observed for a few months before the outburst is closely reminiscent that of V838~Mon \citep[][see Fig. \ref{fig4}, top-left panel]{mun02,gor02,kim02,cra03,cra05} before the main peak.  
We also note that the absolute magnitude of AT~2019zhd, both before and during the outburst, 
is intermediate between luminous extra-galactic events such as SNhunt248  and AT~2015dl, 
and the faint LRNe observed in the Milky Way, such as V1309~Sco  and OGLE-2002-BLG-360.  
Hence, its absolute magnitude at peak ($M_r=-9.61\pm0.08$ mag) is somewhat comparable with that of V838~Mon.

Fig. \ref{fig4} (bottom-left panel) shows that the $B-V$ colour evolution of AT~2019zhd during the first month after the outburst ranges from about 0 (at the blue peak)
to 2 mag at the end of the plateau. This colour becomes  redder
much more rapidly than what has been observed in the extra-Galactic LRNe discussed in \citet{pasto19a}, and this evolution is comparable to those shown by fainter Galactic objects, 
such as V1309~Sco.

Figure \ref{fig4} (right panel) compares the quasi-bolometric evolution of AT~2019zhd, obtained after integrating the flux over the optical domain only, with those
of a wide sample of LRNe. The luminosity of AT~2019zhd at peak, $1.4 \times 10^{39}$ erg s$^{-1}$, is similar to that of the prototypical V838~Mon.  
The optical luminosity of AT~2019zhd is over two orders of magnitude fainter than that of the luminous SNhunt248 \citep{kan15}, AT~2014ej \citep{str20}, and AT~2017jfs \citep{pasto19b}, 
and about 25 times brighter than V1309~Sco. After maximum, the light curve decline is initially 
very fast, it flattens after one week, and increases again at late phases. The figure also shows that fainter LRNe have light curves
with shorter duration than those of the most luminous counterparts. 
In general, AT~2019zhd has a luminosity similar to V838~Mon and the other two LRNe in M~31, although with a faster-evolving light curve, which is reminiscent of that of V1309~Sco.
The short-duration plateau of AT~2019zhd is likely due to a smaller ejected mass than other LRNe.

   \begin{figure}
   \centering
   \includegraphics[width=8.8cm,angle=0]{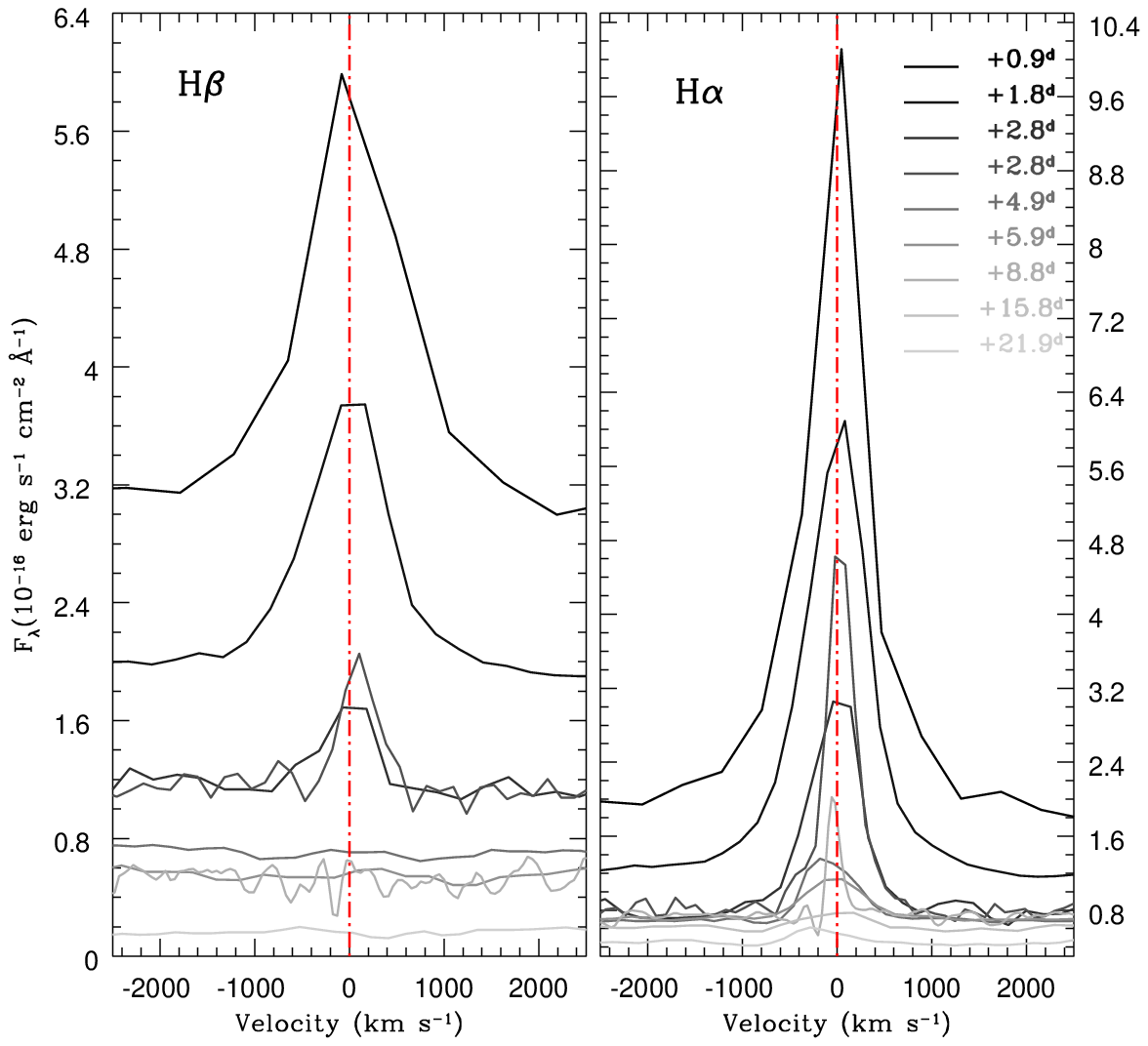}
   \caption{Evolution of the H$\beta$ ({\it left}) and H$\alpha$ ({\it right}) profiles in the spectra of AT~2019zhd.
The adopted redshift, $z=-0.001764$, it that of the emission line source 2035 of \citet[][see Sect. \ref{host}]{mer06},
close to the LRN site.}
              \label{fig6}
    \end{figure}

   \begin{figure}
   \centering
   \includegraphics[width=8.8cm,angle=0]{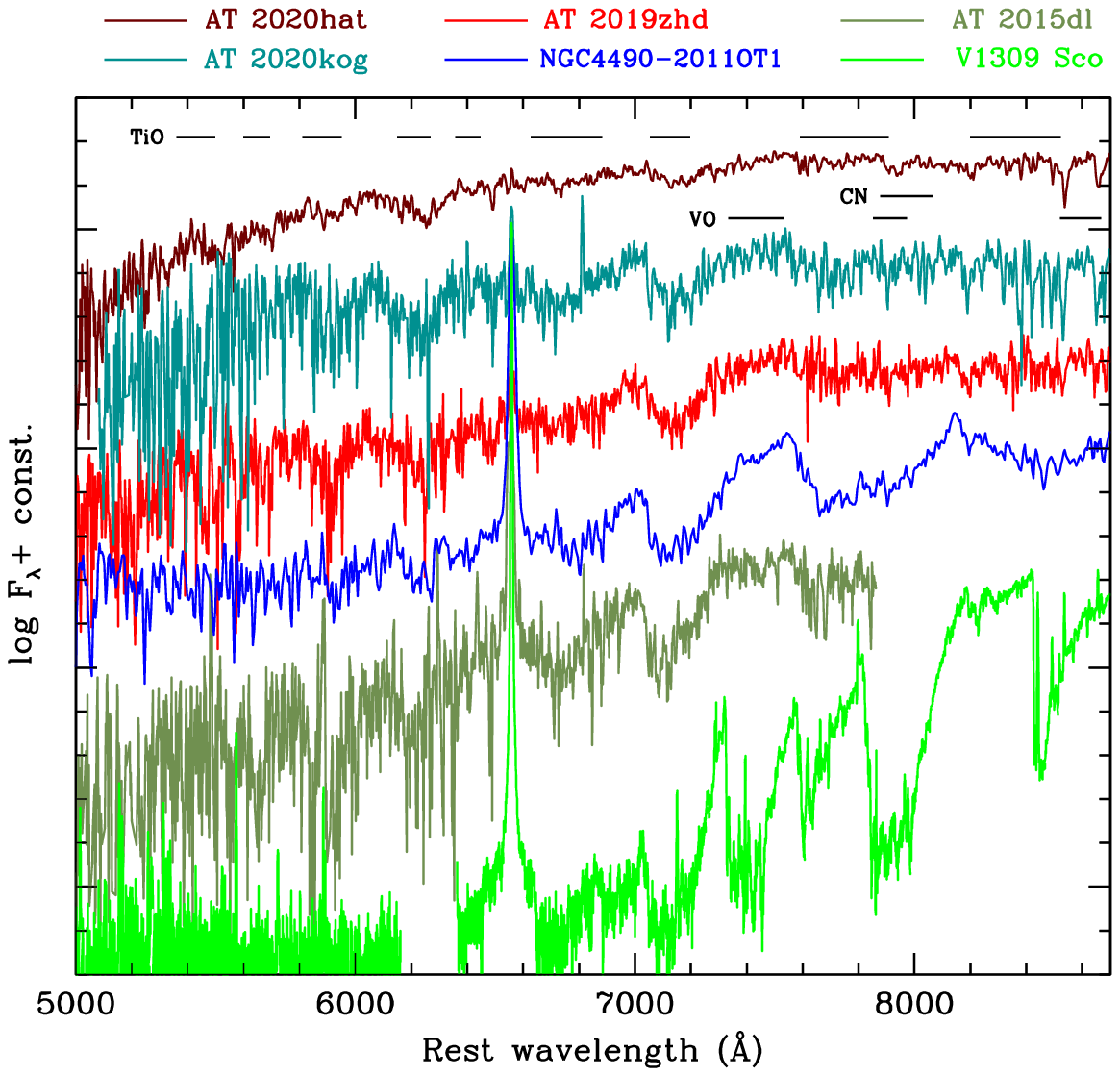}
   \caption{Comparison of late-time spectra of LRNe, with the most prominent molecular bands highlighted. The sample includes AT~2019zhd (phase $\sim29$~d), AT~2020hat  ($\sim76$~d) and AT~2020kog \protect\citep[$\sim100$~d;][]{pasto20}, NGC4490-2011OT1 \protect\citep[$\sim218$~d;][]{pasto19a}, AT~2015dl \protect\citep[$\sim252$~d;][]{bla17}, and V1309~Sco \protect\citep[$\sim62$~d;][]{mas10}. }
              \label{fig7}
    \end{figure}

\section{Spectral evolution} \label{spectroscopy}

The spectroscopic follow-up of AT~2019zhd lasted about one month. The monitoring campaign was limited by the short visibility window of the object, 
which was observable only soon after sunset. Nonetheless, we obtained ten spectra using the 2.0~m Liverpool Telescope equipped with SPRAT \citep{pia14},
the 2.56~m NOT with ALFOSC, the 10.4~m Gran Telescopio Canarias (GTC) with OSIRIS, the 1.82~m Copernico Telescope with AFOSC, and the 1.22~m Galilei Telescope with a B$\&$C spectrograph.
We monitored all crucial phases of the LRN evolution, from the initial blue peak to the end of the plateau. The spectroscopic data were reduced through
standard tasks in {\sc IRAF}\footnote{{\sc IRAF} is distributed by the National Optical Astronomy Observatory, which is operated by the Association of Universities for Research in Astronomy (AURA) under a cooperative agreement with the National Science Foundation.} or dedicated pipelines, such as the {\sc FOSCGUI}\footnote{{\sc FOSCGUI} is a graphic user interface developed by E. Cappellaro, and aimed 
at extracting supernova spectroscopy and photometry obtained with FOSC-like instruments. A package description can be found at \url{http://sngroup.oapd.inaf.it/foscgui.html}.}.
Technical information on the spectra is reported in Table \ref{tab1}, while the entire spectral sequence is shown in Fig. \ref{fig5}.

Early spectra, taken soon after the blue peak, show a blue continuum with superposed prominent emission lines of the Balmer series. A number of Fe~II features
with a predominant emission component are also clearly detected, along with Ca~II H$\&$K, which is instead observed in absorption. The continuum temperature decreases
from $T =  9500 \pm 700$ to $7900 \pm 500$~K from 0.9 d to 2.8~d after maximum. The best resolution B$\&$C spectrum at +2.8~d \citep{mun20} allows us to constrain an upper limit 
for the full-width at half-maximum (FWHM) velocity for the H$\alpha$ emission, which is about 280 km~s$^{-1}$.

The following spectra show a remarkable transition, with the continuum becoming rapidly redder ($T = 5400 \pm 400$ K at 4.9~d, $T =  4700 \pm 400$ K at 5.9~d,
and $T = 4100  \pm 600$ K at 15.8~d). In a moderate resolution GTC spectrum obtained at 8.8~d after maximum, H$\alpha$ shows a P-Cygni profile with an
absorption component which is blue-shifted by about 160 km~s$^{-1}$. 
During the first two weeks past maximum, the Balmer emission components become progressively weaker, and only H$\alpha$ is barely visible at 15.8~d, 
with a slightly blueshifted peak (see Fig. \ref{fig6}). In this phase, the spectrum is dominated by a forest of narrow absorption metal lines. 
Following \citet[][see their figure 10]{pasto19a}, we identify the prominent Fe~II multiplets in absorption, along with Sc~II, Ba~II, Ti~II, Na~I doublet, Ca~II and O~I.
This transitional spectrum is typical of LRNe during the second, redder peak \citep{smi16,bla17,pasto19b,cai19,str20}, although here the transition is much 
more rapid, in agreement with the rapid light curve evolution.

When the light curve starts the fast post-plateau decline, the spectral continuum temperature further decreases
to  $T=3100\pm400$ K at 21.9~d, and  $T=2600\pm500$ K at 28.9~d. In these later spectra, broad absorption bands produced by molecules are clearly detected at about 5450~\AA, 5900~\AA, 6250~\AA,
6750~\AA, 7150~\AA, 7700~\AA~and 8300~\AA; in particular, TiO bands are becoming prominent \citep{val98}. Again, the presence of molecular bands is a common feature
of late-time optical spectra of LRNe \citep{smi16,bla17,pasto19a,pasto19b,cai19}, although it happens earlier in faint, 
fast-evolving LRNe observed in the Milky Way \citep[e.g.][]{mar99,mun07,gor07,kam09,mas10,bar14}.
A comparison of the latest spectrum of AT~2019zhd with late spectra of other LRNe is shown in Fig. \ref{fig7}.

   \begin{figure*}
   \centering
   \includegraphics[scale=0.65,angle=270]{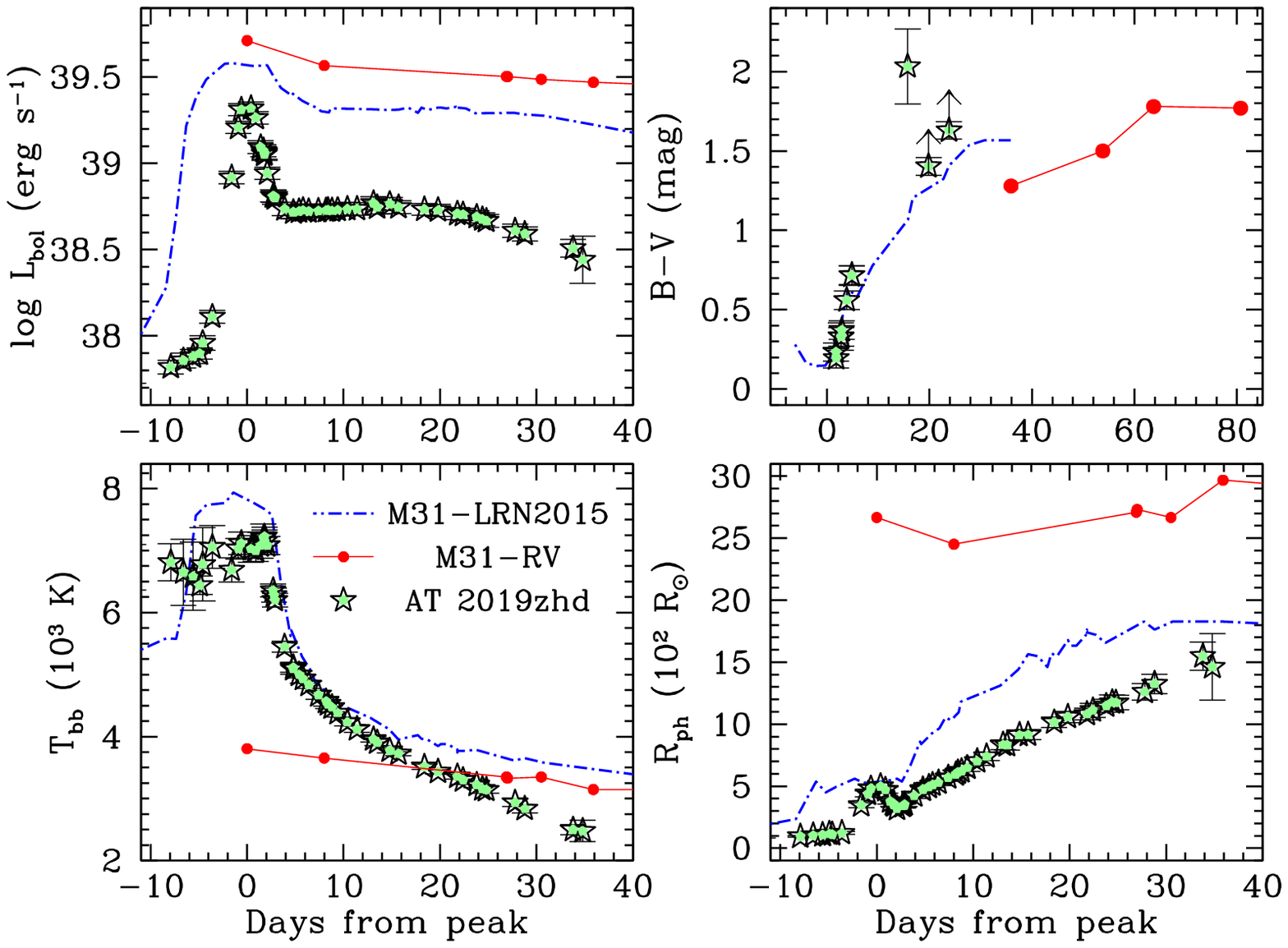}
   \caption{Comparison of the physical parameters of the three LRNe in M~31 (adopting $\mu= 24.47 \pm 0.08$ mag for M~31): AT~2019zhd, M31-LRN2015 and M31-RV. {\it Top-left}: Bolometric light curves. {\it Top-right}: $B-V$ colour evolution.
{\it Bottom-left}: Evolution of the blackbody temperature evolution. {\it Bottom-right}: Evolution of the radius. The data of M31-RV are from \protect\citet{bos04}, those of M31-LRN2015 are from
 \protect\citet{kur15,wil15,lip17,bla20}. For the total reddening of the comparison objects, we adopt $E(B-V) = 0.37$ mag for M31-LRN2015 \protect\citep[from][]{kur15}, and $E(B-V) = 0.12$ mag for M31-RV \protect\citep[][]{bos04}.
}
              \label{fig8}
    \end{figure*}

\section{Bolometric light curve and evolution of the temperature and the radius} \label{radius}

As noted in Sect. \ref{photometry_LRN} (see Fig. \ref{fig4}, right panel), AT~2019zhd belongs to the LRN sample having intermediate luminosity at maximum 
($M_V\sim -10$ mag). This group also includes V838~Mon, and two other LRNe discovered in M~31: M31-LRN2015 \citep{kur15,wil15,lip17,bla20} 
and M31-RV \citep{ric89,mou90,bry92,tom92,bos04}. The fact that only intermediate-brightness LRNe have been discovered in M~31 is somewhat puzzling. While 
 it is not surprising that very luminous LRN events have not been observed so far in M~31 as they are
intrinsically rare \citep{koc14}, faint events similar to V1309~Sco are expected to be common, and they should have been discovered in M~31. 
One may invoke selection effects to explain the lack of faint red nova discoveries in this galaxy. Objects similar to V1309~Sco, at the distance of M~31, 
would have a peak apparent magnitude of about 18, hence they should be comfortably observed, unless they occur
in crowded or dusty regions. 
In this respect, we note that a few tens of nova candidates are discovered every year in M~31 in the magnitude range 17-21 mag, and a large fraction of them 
\citep[nearly 50 $\%$, from the on-line database of M~31 novae\footnote{\url{https://www.mpe.mpg.de/~m31novae/opt/m31}}; see][]{pie10}
remain unclassified. Hence, it is plausible that low-luminosity events similar to V1309~Sco can be found among these unclassified candidates.

\begin{figure*}
	\includegraphics[width=\textwidth]{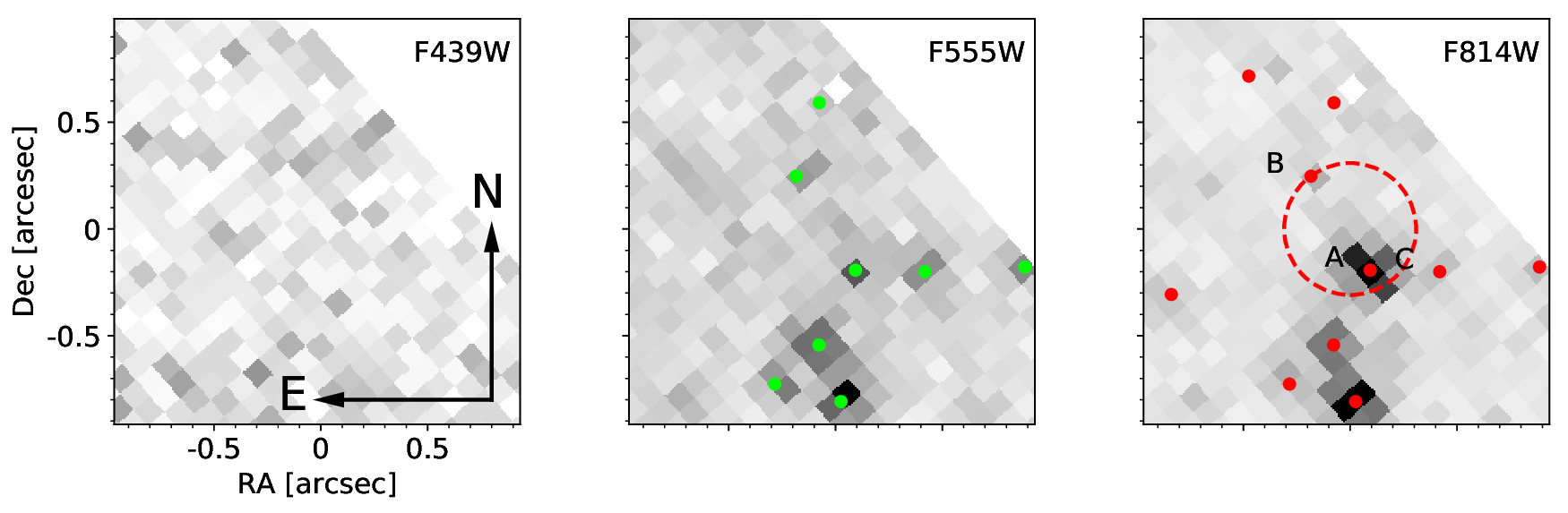}
    \caption{2\arcsec $\times$2\arcsec cutouts from the 1997 HST+WFPC2 images covering the site of AT~2019zhd. The 0.31\arcsec positional uncertainty on AT~2019zhd is indicated with a red dashed circle. Sources detected by {\sc dolphot} at $>5\sigma$ significance in each filter are marked with coloured points. Sources A, B and C as discussed in the text are labelled.}
    \label{fig:progenitor}
\end{figure*}

To best characterise the three LRNe in M~31, we now compare their parameters inferred from the available photometric data, 
adopting a similar approach as \citet[][see their figure 2]{mac17}. For M31-LRN2015, we adopt a moderate-reddening scenario \citep[$E(B-V) = 0.37$ mag,][]{kur15}.

In  Fig. \ref{fig8} (top-right panel), we show the  $B-V$ colour curve as in Fig. \ref{fig4} (bottom-left), but we limit the sample to the three LRNe in M~31. The colour evolution of the three 
objects is remarkably similar, with $B-V$ growing from about 0.1 mag at maximum to over 1.4-2 mag (depending on the object) after one month. 
The $B-V$ colour increases more rapidly in AT~2019zhd than in the other two LRNe, reaching $B-V\sim2$ mag at about 15 d after maximum. 
The late evolution, up to +82~d, is observed only for M31-RV, and the  colour seems to flatten to $B-V \approx 1.8$ mag.

In order to determine the evolution of the bolometric luminosity ($L_{bol}$), the temperature ($T_{bb}$) and the photospheric radius ($R_{ph}$), 
we first construct the spectral energy distribution (SED) for each object, and for any epochs with available multi-band, reddening-corrected photometric data.
When the measurement in a determined band is missing, its flux contribution can be estimated through an interpolation using the magnitudes 
at adjacent epochs, or extrapolating the missing magnitude by assuming a constant colour from the closest epoch.
This assumption is very crude, and is critical mainly in the pre-maximum phases, when the object is expected to have a very rapid colour evolution.

The total flux ($F_{bb}$) and $T_{bb}$ are estimated for each epoch through a Monte Carlo simulation.
Firstly, flux values at different wavelengths ($F_{\lambda}$) are generated with a Gaussian distribution centred at the measured $F_{\lambda}$, and with a standard deviation equal to the measured F$_{\lambda}$ error.
Then, we fitted a black-body to the randomly generated $F_{\lambda,i}$ values, obtaining an i-th estimate of the total flux $F_{bb,i}$ and the temperature T$_{bb,i}$.

After reiterating this procedure for 200 times, we adopt the median values of $T_{bb,i}$ and $F_{bb,i}$ as best estimates of $F_{bb}$ and $T_{bb}$ for that epoch, and the errors are given by the standard deviation
of the 200 $T_{bb,i}$ and $F_{bb,i}$ estimates. Adopting the distance given in Sect. \ref{host} and assuming spherical symmetry for the emitting source, from  $F_{bb}$ we infer $L_{bol}$ for that epoch.
Finally, $R_{ph}$ corresponing to the above $L_{bol}$ is obtained through the Stefan-Boltzmann law for an emitting sphere. Trivial error propagation provides the uncertainties on the $L_{bol}$ and $R_{ph}$ estimates.
The evolution of $L_{bol}$, $T_{bb}$ and $R_{ph}$ are obtained by repeating the full procedure for all available epochs. We remark that, while in general a temperature inferred through
black-body fits to the SED is not a good proxy for the photospheric temperature in spectra dominated by deep TiO features \citep[e.g.][]{flu94}, this is acceptable in the case of AT~2019zhd,
as the TiO bands become relatively prominent only at late epochs ($\sim$30 days after maximum, see Sect. \ref{spectroscopy}).

The resulting bolometric light curves of the three objects are shown in Fig. \ref{fig8} (top-left panel). 
AT~2019zhd peaks at $L\approx2.1\times10^{39}$ erg s$^{-1}$ (hence, $L\sim5.5\times10^5$ L$_\odot$), about two times fainter than M31-LRN2015 ($L\approx4\times10^{39}$ erg s$^{-1}$).
We note that the bolometric light curve of M31-RV\footnote{We assume the epoch of the early peak is MJD = 47355.5 \protect\citep{mou90}. Unfortunately,  no precise colour information is available at that epoch, 
and this uncertainty affects the accuracy of the bolometric correction. The first epoch with multi-band photometry was obtained at about 36~d days after the peak.} shown here is the most luminous in the M~31 LRN sample, peaking at $L\approx5.1\times10^{39}$ erg s$^{-1}$.
This is due to  the large NIR contribution to the bolometric luminosity, obtained through the NIR observations originally presented by \citet{mou90}.

The evolution of $T_{bb}$ is shown in Fig. \ref{fig8} (bottom-left panel). While the pre-maximum evolution
is very uncertain for all objects due to the incomplete colour information, from the available data we note that M31-LRN2015 peaks at almost 8000~K,
AT~2019zhd peaks at about 7000~K, while M31-RV seems to peak at much lower temperatures ($T_{bb} \approx 3800$~K).
Soon after maximum, their temperatures undergo a fast decline during the first week, and then the decline rates become slower. At one month after maximum, 
the temperature measured for AT~2019zhd was $T_{bb} \approx 2500$~K, while both  M31-RV and M31-LRN2015 have hotter blackbody temperatures, i.e.
 $T_{bb} \approx 3200$~K and  3600~K, respectively.

Finally, Fig. \ref{fig8} (bottom-right panel) shows the evolution of $R_{ph}$. Again, the  evolution of $R_{ph}$ is very uncertain in the 
pre-peak phases for the reasons mentioned above. At maximum,  $R_{ph} \sim 500~R_\odot$ in both AT~2019zhd and M31-LRN2015. After a short-lasting decline soon after peak,  
$R_{ph}$ increases almost linearly in AT~2019zhd until phase $\sim$35~d ($R_{ph} \sim 1500~R_\odot$). At the same epoch, $R_{ph} \sim 1800~R_\odot$ in M31-LRN2015, while
M31-RV appears  to be much more expanded, with $R_{ph} \sim 3000~R_\odot$.

We note that in AT~2019zhd and M31-LRN2015 the evolution of $T_{bb}$ and $R_{ph}$ follows a similar trend as more luminous LRNe, such as AT~2020kog \citep{pasto20}, 
AT~2018hso \citep{cai19}, AT~2017jfs \citep{pasto19b}, and AT~2014ej \citep{str20}, that have higher temperature ($T_{bb} > 7000$~K) and a more 
compact radius at the early maximum. In contrast, M31-RV is much redder (the temperature is lower by a factor of two) and has a larger radius at maximum. 
This behaviour is reminiscent of those of AT~2015dl \citep{bla17} and AT~2020hat \citep{pasto20}, which do not show a prominent early, hot peak.
The variety in early-time temperatures and radii of LRNe could be a consequence of the different geometry of the expelled gas, the radii of the two stellar components, 
the total masses involved, and the final outcome of the dynamic interaction process.
Detailed modelling is thus necessary to infer whether the heterogeneous evolution of the above parameters brings some information on the fate of the system.
In particular, the LRN observables are expected to be different when the post common-envelope evolution 
has led to coalescence, or eventually to a new stable binary configuration.

\section{The quiescent progenitor} \label{progenitor}

\begin{figure}
	\includegraphics[width=\columnwidth]{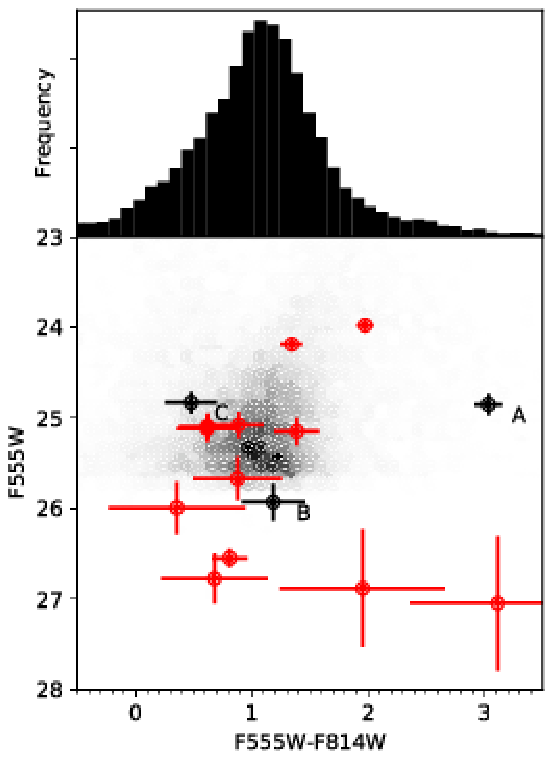}
    \caption{{\it Top}: histogram of the $F555W-F814W$ colour for all sources on the WF4 chip, showing how Source A is clearly redder than the vast majority of these. {\it Bottom}:
colour-magnitude diagram of the sources found by {\sc dolphot} in the vicinity of AT~2019zhd. The grey density plot shows all sources detected at $>5\sigma$ on the WF4 chip, 
while the red points show sources within the 2\arcsec$\times$2\arcsec region shown in Fig. \ref{fig:progenitor}. Sources A, B and C as discussed in the text are labelled.}
    \label{fig:progenitor_CMD}
\end{figure}

In order to characterise the progenitor of AT 2019zhd, we searched in the Mikulski Archive for Space Telescopes for Hubble Space Telescope (HST) images covering the site of the transient. While M~31 has been observed extensively with HST, the site of AT~2019zhd was only covered by one epoch of multi-band imaging, due to its small field of view. These images were taken with WFPC2 on 1997 December 9, with exposure times of 2$\times$350~s,  2$\times$260~s and 2$\times$260~s in the $F439W$, $F555W$ and $F814W$ filters respectively. The site of AT~2019zhd falls on the WF4 chip, which has a pixel scale of 0.1\arcsec~pix$^{-1}$, and is unfortunately $<$10 pixels from the edge of the detector.

To precisely localise the position of AT~2019zhd on the WFPC2 images, we performed differential astrometry between the $F814W$ images and a 35~s $r$-band image of the transient taken on 2020 February 17 with the NOT plus ALFOSC. Seven point sources common to both images were used to determine a geometric transformation. From this, we find the pixel coordinates of AT~2019zhd to be (792.67, 522.15) on the $F814W$ image. The associated uncertainty on this position is 103 mas ($\sim$1 WF pixel) from the rms scatter of the sources used in deriving the transformation. However, due to systematic effects, this is likely an underestimate of the true uncertainty on the position. Firstly, the only sources visible in common to both the NOT and HST images were bright foreground stars. These will potentially exhibit proper motion at $\gtrsim$mas level over the 22 years between the HST and NOT images. As AT~2019zhd lies at the extreme edge of the WF4 chip we must extrapolate the geometric transformation (since the reference sources used for the alignment all lie to one side of the transient). Finally, our geometric transformation has six free parameters, while we are only using seven sources to derive the transformation. All things considered, we adopt three times the transform uncertainty (i.e. 0.31\arcsec) as a conservative uncertainty on the location of AT 2019zhd in the HST images.

In order to identify sources on the HST images, we used the {\sc dolphot} photometry package \citep{Dolp00}. All sources detected by {\sc dolphot} at $>3\sigma$ confidence are plotted in Fig. \ref{fig:progenitor}. One source was found to be consistent with the position of AT~2019zhd (designated as Source A), along with two sources just outside our adopted positional uncertainty (Sources B and C). While we regard Source A as a credible counterpart to AT~2019zhd on its location, we note that it also has markedly different colour with respect to the other sources in the field. PSF-fitting photometry was performed for Source A using {\sc dolphot}, and we find magnitudes for the source (in the Vegamag system) of $F555W=24.85\pm0.12$ and $F814W=21.82\pm0.03$ mag, while it is not detected in the $F439$ filter, down to $\sim25.6$ mag.
Source A is extremely red, hence it is not surprising  it is undetected in the {\it F439W} filter.
A comparison to other sources in the field in Fig. \ref{fig:progenitor_CMD} shows it to be far redder than the majority of sources in the field, and this difference with the local stellar population makes source A as an appealing progenitor candidate for AT~2019zhd.

For Source B, we measure $F555W=25.93 \pm0.12$ mag; $F814W=24.74\pm0.18$ mag; while for Source C, we find $F555W=24.83 \pm0.12$ mag; $F814W=24.36\pm0.19$ mag. As these are further from the location of AT~2019zhd, and appear very similar to other sources in the field, we regard them as unlikely to be the progenitor.

Turning back to Source A, after correcting for foreground reddening (E($B-V$)$=0.055$ mag), and adopting a distance modulus for M~31 of $\mu=24.47\pm0.08$ mag, we find an absolute magnitude of M$_{F555W}= 0.21\pm0.14$ mag and a $F555W-F814W$ colour of $2.96\pm0.12$~mag. While this candidate progenitor is much redder than those of most LRNe studied so far \citep[e.g.][]{pasto19a}, its red colour is somewhat reminiscent of that of the progenitor of AT~2020hat \citep{pasto20}. The progenitor of M31-LRN2015 has been also detected in archive images, showing a brighter absolute magnitude (M$_V= -1.50\pm0.23$ mag) and a bluer colour
\citep[$V-I = 1.05 \pm 0.15$ mag;][]{wil15}.
We remark that, as most (if not all) LRNe are closely interacting binary systems, the two stellar components contribute to the magnitudes and the colours inferred here.


An alternative scenario is that none of the three sources mentioned above is the progenitor of AT~2019zhd, hence source A 
is merely a chance alignment with the location of the LRN.
Accounting for the detection limits of the HST images, a non-detected progenitor would have $F439W > 25.6$ mag, $F555W > 25.7$ mag and $F814W > 25.4$ mag, hence $M_{F439W} > 0.9$ mag, $M_{F555W} > 1.1$ mag and $M_{F814W} > 0.8$ mag.
Such a faint source would not comfortably match the correlations between progenitor versus light curve absolute magnitudes \citep{pasto19a}, as it would
fainter than the expected absolute magnitude of the progenitor of a LRN whose light curve peaks at $M_V \approx -9.1$ mag \citep[see, also, figure 9 of][]{pasto20}.
Moreover, the exceptionally red colour of Source A is markedly different from other sources in the field, and may be explained by a puffed-up star that has formed a common-envelope.
For this reason, while we believe that source A remains the most plausible progenitor of AT~2019zhd, a bluer progenitor below the detection threshold of the HST images 
cannot be definitely ruled out.

\section{Discussion and conclusions} \label{discussion}

In order to  infer the $V$-band magnitudes of AT~2019zhd at a few critical phases of its evolution, we apply the conversion relations between photometric systems from \citet{jor06}.
During the slow magnitude rise before the LRN outburst, AT~2019zhd reaches a maximum $V$-band absolute magnitude $M_V=-5.74\pm0.28$ mag. Superposed to the global rise, the
light curve shows some fluctuations, similar to those observed for instance in M31-LRN2015. 

As discussed by \citet{pej17}, the dynamical mass transfer in a binary system, which happens when one star has filled its Roche lobe, may cause a systemic mass loss with 
a double effect: the decrease of the angular momentum  (which may eventually lead the binary to merge) and the formation of an optically thick expanding 
cocoon (the CE) that engulfs the stellar system \citep{pac76}. This phase is characterised by a substantial luminosity decline. 
Shocks between material ejected in more recent mass-loss events and the CE may cause a renewed luminosity rise, as observed before the outburst of M31-2015LRN and other LRNe  
\citep{bla20}. The non-monotonic light curve of AT~2019zhd before the main outburst (see Fig. \ref{fig2}) can be explained with a similar mechanism as proposed for M31-2015LRN.

The subsequent luminous blue peak in the light curve of AT~2019zhd at $M_V=-9.08\pm0.12$ mag is produced by a violent gas ejection during the merging process,
which then interacts with the pre-existing circumstellar material \citep{mac17}.
The average magnitude of the plateau (which is considered analogous of the red peak in more luminous events) is $M_V=-7.59\pm0.32$ mag. 
This feature in the LRN light curve, characterised by a roughly constant effective temperature, has been explained as due to a hydrogen recombination front 
which propagates through the expanding gas \citep{iva13,lip17}, similar to that observed during the plateau phase in type IIP supernovae \citep{pop93}. 
This scenario is also confirmed by the Balmer decrement, which ranges from 2.4 to 2.8 in the first two spectra, consistent with the expectations for Case B 
recombination, along with the decrease of the Balmer emission lines intensity.

While some observational properties (e.g. the light curve shape with an early, sharp peak followed by a broader maximum or a plateau, and the late spectra
reminiscent of those of intermediate-to-late M spectral types) are shared among all LRNe, we have seen in Sect. \ref{radius} that other parameters (peak 
luminosity, effective temperature, and photospheric radius) show a wide diversity which could be due to the dynamical processes occurring in the binary system, and the physical properties 
of the individual stellar components. Exploring the diversity in the observed parameters through hydro-dynamical models can give some constraints on the main parameters of the
binary system and, eventually, the final merger.

\citet{koc14} proposed a tight relation between the luminosity of the outburst and the total mass of the binary. This is also supported by the correlations of the LRN luminosity
with that of the progenitor system before the outburst \citep{pasto19a}, and with the  outburst duration \citep{pasto20}.
Objects with L$_{bol}$ at maximum exceeding $10^{41}$ erg s$^{-1}$ have been discovered, along with transients whose L$_{bol}$ is 3-4 orders of magnitude smaller. 
In the luminosity distribution of LRNe, AT~2019zhd lies in the middle \citep[see figure 9 in][]{pasto20}, sharing similar photometric parameters with objects such as 
V838~Mon \citep[e.g.][]{gor02,kim02,mun02,cra03,gor04,cra05,gor20}, M31-RV \citep{bon03,bos04}, and M31-LRN2015 \citep{kur15,wil15,lip17,bla20}. This implies that 
these transients were likely produced by stellar systems with relatively similar masses, intermediate between those of the two groups mentioned above.

The system leading to the M31-LRN2015 event was observed over a decade before the LRN outburst in HST images at
$M_V=-1.50\pm0.23$ mag and $V-I=1.05\pm0.15$ mag \citep{wil15} \citep[see, also,][]{dong15}, which are typical of a late-G to early K-type stars.
The progenitor of M31-LRN2015 was somewhat redder than those discussed in \citet{pasto19a}. 
It is important to note, however, that \citet{mac17} estimated a somewhat bluer intrinsic colour adopting a higher reddening scenario (up to $V-I\sim0.6$ mag,
giving an F spectral type source, see their figure 4).
The M31-LRN2015 precursor was likely formed by a primary sub-giant of 3-5.5 M$_\odot$ with a radius of $\sim$35 R$_\odot$, 
and a lower-mass (0.1-0.6 M$_\odot$) main sequence  companion \citep{mac17,lip17}. 
The M31-RV LRN event has somewhat similar photometric characteristics, and the progenitor belongs to a population 
of old red giants in the M~31 halo. While no information is available on the system before the M31-RV outburst, an
in-depth analysis of the LRN location  years after the outbursts  was performed by some authors, 
although the results do not give firm constraints on the identification of the survivor. Most likely, the merger 
faded below the detection limit due to dust obscuration, or the survivor is eventually one of the red giants 
in the field\footnote{We note that this conclusion was questioned by \citet[][]{sha10}.} \citep{bon06,bon11,bon18}.
In the case of AT~2019zhd, a stellar source at the expected position of the quiescent progenitor was tentatively identified in pre-explosion HST archive images,
with absolute magnitude $M_{F555W}=0.21\pm0.14$ mag, and a reddening-corrected colour $F555W-F814W=2.96\pm0.12$ mag, which is consistent with an intermediate M-type source.

In this context, while the V838~Mon LRN event was relatively similar to the above objects, its progenitor system was more massive and bluer, and hosted 
in a young stellar population environment \citep[][and references therein]{bar17,gor20}, somewhat more similar to the environments of the brightest extra-galactic LRNe 
\citep{smi16,mau15,gor16,bla17,pasto19a,cai19}.

Recent studies on LRNe have confirmed a large heterogeneity in their properties, which is expected since binary interaction leading to merging events
may involve stars spanning an enormous range of masses. Further observational efforts and investments in developing theoretical models are necessary to 
provide robust correlations between the observational parameters and those of the binary progenitor system.

\begin{acknowledgements}

We acknowledge with thanks the variable star observations from the AAVSO International Database contributed by observers worldwide and used in this research.\\

MF gratefully acknowledges the support of a Royal Society – Science Foundation Ireland University Research Fellowship.
EK and MS are supported by generous grants from  Villum FONDEN (13261,28021) and by a project grant (8021-00170B) from the Independent Research Fund Denmark.
SJS and KWS acknowledge funding from STFC Grants ST/P000312/1, ST/T000198/1 and ST/S006109/1.
DJ acknowledges support from the State Research Agency (AEI) of the Spanish
Ministry of Science, Innovation and Universities (MCIU) and the European Regional
Development Fund (FEDER) under grant AYA2017-83383-P. DJ also acknowledges
support under grant P/308614 financed by funds transferred from the
Spanish Ministry of Science, Innovation and Universities, charged to the General
State Budgets and with funds transferred from the General Budgets of the
Autonomous Community of the Canary Islands by the Ministry of Economy, Industry,
Trade and Knowledge.\\

This research is based on observations made with the Nordic Optical Telescope, operated by the Nordic Optical Telescope Scientific Association at the Observatorio del Roque de los Muchachos, La Palma, Spain, 
of the Instituto de Astrofisica de Canarias; the Gran Telescopio Canarias, installed at the Spanish Observatorio del Roque de los Muchachos of the Instituto de Astrofísica de Canarias, in the island of La Palma;
the Liverpool Telescope operated on the island of La Palma by Liverpool John Moores University at the 
Spanish Observatorio del Roque de los Muchachos of the Instituto de Astrof\'isica de Canarias with financial support from the UK Science and Technology Facilities Council; 
the 1.82~m Copernico Telescope and the 67/92~cm Schmidt Telescope of INAF-Osservatorio Astronomico di Padova at Mt. Ekar; and the 1.22~Galileo Galilei Telescope of the Padova University in the Asiago site.
 \\

This work has made use of data from the Asteroid Terrestrial-impact Last Alert System (ATLAS) project. ATLAS is primarily funded to search for near earth asteroids through NASA grants NN12AR55G, 80NSSC18K0284, and 80NSSC18K1575; byproducts of the NEO search include images and catalogues from the survey area. The ATLAS science products have been made possible through the contributions of the University of Hawaii Institute for Astronomy, the Queen's University Belfast, the Space Telescope Science Institute, and the South African Astronomical Observatory, and The Millennium Institute of Astrophysics (MAS), Chile.\\

The Pan-STARRS1 Surveys (PS1) and the PS1 public science archive have been made possible through contributions by the Institute for Astronomy, the University of Hawaii, the Pan-STARRS Project Office, the Max-Planck Society and its participating institutes, the Max Planck Institute for Astronomy, Heidelberg and the Max Planck Institute for Extraterrestrial Physics, Garching, The Johns Hopkins University, Durham University, the University of Edinburgh, the Queen's University Belfast, the Harvard-Smithsonian Center for Astrophysics, the Las Cumbres Observatory Global Telescope Network Incorporated, the National Central University of Taiwan, the Space Telescope Science Institute, the National Aeronautics and Space Administration under Grant No. NNX08AR22G issued through the Planetary Science Division of the NASA Science Mission Directorate, the National Science Foundation Grant No. AST-1238877, the University of Maryland, Eotvos Lorand University (ELTE), the Los Alamos National Laboratory, and the Gordon and Betty Moore Foundation.\\

Lasair is supported by the UKRI Science and Technology Facilities Council and is a collaboration between the University of Edinburgh (grant ST/N002512/1) and Queen’s University Belfast (grant ST/N002520/1) within the LSST:UK Science Consortium. 
This publication is partially based on observations obtained with the Samuel Oschin 48-inch Telescope at the Palomar Observatory as part of the Zwicky Transient Facility project. ZTF is supported by the National Science Foundation grant No. AST-1440341 and a collaboration including Caltech, IPAC, the Weizmann Institute for Science, the Oskar Klein Center at Stockholm University, the University of Maryland, the University of Washington, Deutsches Elektronen-Synchrotron and Humboldt University, Los Alamos National Laboratories, the TANGO Consortium of Taiwan, the University of Wisconsin at Milwaukee, and Lawrence Berkeley National Laboratories. Operations are conducted by COO, IPAC, and UW. This research has made use of ``Aladin sky atlas'' developed at CDS, Strasbourg Observatory, France.\\

This work made use of the NASA/IPAC Extragalactic Database (NED), which is operated by the Jet Propulsion Laboratory, California Institute of Technology, under contract 
with NASA. We also used NASA's Astrophysics Data System. \\

\end{acknowledgements}

%
%

\begin{appendix}
\onecolumn

\section{Supplementary material}

\begin{table}[h]
\caption{ Johnson-Bessell $B,V$ photometry of AT~2019zhd and associated errors (in Vega magnitude scale).}\label{tabA1}
     $$         \begin{array}{ccccc}
\hline \hline
            \noalign{\smallskip}
Date &  MJD &  B & V & Instrument \\ 
\hline
2019-12-17  &   58834.19 &   $>$19.58       &  20.114~  (0.296)  &   0  \\  
2020-02-14  &   58893.18 &        --        &  15.56~   (0.10)   &   1  \\  
2020-02-14  &	58893.73 &        --        &  15.806~  (0.032)  &   2  \\  
2020-02-14  &	58893.74 &        --        &  15.809~  (0.047)  &   2  \\  
2020-02-14  &   58893.75 &  16.185~ (0.011) &  15.896~  (0.011)  &   3  \\  
2020-02-14  & 	58893.80 &  16.142~ (0.016) &  15.893~  (0.019)  &   4  \\  
2020-02-14  &	58893.85 &        --        &  15.953~  (0.026)  &   2  \\  
2020-02-14  &	58893.86 &        --        &  15.963~  (0.020)  &   2  \\  
2020-02-14  &	58893.86 &        --        &  15.965~  (0.020)  &   2  \\  
2020-02-14  &	58893.88 &        --        &  15.953~  (0.029)  &   2  \\  
2020-02-14  &	58893.88 &        --        &  15.986~  (0.028)  &   2  \\  
2020-02-15  & 	58894.74 &  16.822~ (0.049) &  16.437~  (0.044)  &   5  \\  
2020-02-15  & 	58894.79 &  16.824~ (0.014) &  16.440~  (0.012)  &   4  \\  
2020-02-15  &	58894.82 &        --        &  16.446~  (0.048)  &   2  \\  
2020-02-15  &	58894.84 &        --        &  16.434~  (0.034)  &   2  \\  
2020-02-15  &	58894.85 &        --        &  16.429~  (0.039)  &   2  \\  
2020-02-15  &	58894.86 &        --        &  16.424~  (0.037)  &   2  \\  
2020-02-15  &	58894.87 &        --        &  16.457~  (0.035)  &   2  \\  
2020-02-15  &	58894.88 &        --        &  16.453~  (0.042)  &   2  \\  
2020-02-15  &	58894.89 &        --        &  16.463~  (0.036)  &   2  \\  
2020-02-15  & 	58894.87 &  16.898~ (0.015) &  16.472~  (0.011)  &   6  \\  
2020-02-16  &   58895.82 &        --        &  16.629~  (0.033)  &   2  \\  
2020-02-16  & 	58895.86 &  17.320~ (0.015) &  16.705~  (0.011)  &   6  \\  
2020-02-17  &	58896.74 &        --        &  16.694~  (0.036)  &   2  \\  
2020-02-17  & 	58896.83 &  17.547~ (0.022) &  16.776~  (0.016)  &   6  \\  
2020-02-17  &   58896.84 &  17.553~ (0.011) &  16.781~  (0.011)  &   7  \\  
2020-02-18  &   58897.82 &        --        &  16.657~  (0.054)  &   2  \\  
2020-02-19  &	58898.92 &        --        &  16.816~  (0.134)  &   2  \\  
2020-02-22  & 	58901.01 &        --        &  16.942~  (0.057)  &   8  \\  
2020-02-28  & 	58907.75 &  19.444~ (0.227) &  17.357~  (0.028)  &   4  \\  
2020-03-03  & 	58911.83 &    $>$19.27      &  17.814~  (0.079)  &   6  \\  
2020-03-07  & 	58915.77 &        --        &  18.150~  (0.129)  &   5  \\  
2020-03-07  & 	58915.83 &  $>$19.83        &       --           &   6  \\  
2020-03-08  & 	58916.79 &        --        &  18.356~  (0.132)  &   5  \\  
2020-03-16  &   58924.77 &        --        &  20.055~  (0.336)  &   5  \\  
\hline                                                                                                                   
         \end{array}
     $$ 
                                                                                                                     
\tablefoot{                                                                                                              
0 = 2.0m Faulkees North Telescope + fa05 camera (Hawaii Isl., USA);
1 = CBAT Transient Object Followup Reports (\url{http://www.cbat.eps.harvard.edu/unconf/followups/J00403785+4034529.html});                   
2 = AAVSO Observations from the AAVSO International Database \protect\citep{kaf20};
3 = ANS Collaboration telescopes ID 310 and 2202 \protect\citep{mun20};          
4 = 1.82~m Copernico Telescope + AFOSC (Cima Ekar, Asiago, Italy);                                                   
5 = 67/92~cm Schmidt Telescope + Moravian G4-16000LC + KAF-16803 CCD CCD (Cima Ekar, Asiago, Italy);    
6 = 2.0~m Liverpool Telescope (LT) + IO:O (La Palma, Canary Islands, Spain);                                            
7 = 2.56~m Nordic Optical Telescope (NOT) + ALFOSC (La Palma, Canary Islands, Spain);                                   
8 = 0.61~m Planewave CDK24 telescope + Apogee CG-16M Camera (Burke-Gaffney Observatory, Saint Maryś University, Canada).}                                                                                                                       
\end{table}                                                                                                          

\longtab[2]{
\begin{landscape}
\begin{longtable}{ccccccccc}
\caption{\label{tabA2} Optical photometry of AT~2019zhd: Sloan $u,g,r,i,z$ and  magnitudes in the natural $w$-PanSTARRS and $orange$-ATLAS ($o$)
photometric systems (in column 8). All data are the AB magnitude scale.}\\
\hline \hline
\tiny
Date &  MJD &  $u$ & $g$ & $r$ & $i$ & $z$ & $w/o$ & Inst. \\ 
\hline
\endfirsthead
\caption{continued.}\\
\hline
Date &  MJD &  $u$ & $g$ & $r$ & $i$ & $z$ & $w/o$ & Inst. \\ 
\hline
\endhead
\hline
\endfoot
\endlastfoot
2007-08-13 &	54325.61  & 	      --      &        --      &  $>$20.32      &        --      &        --      &        --      &     1  \\ 
2017-06-17 &    57921.14  &          --       &        --      &        --      &    $>$22.38    &        --      &       --       &     2  \\ 
2017-12-25 &    58122.24  &          --       &        --      &        --      &    $>$21.84    &        --      &       --       &     3  \\ 
2018-12-12 &    58464.23  &          --       &        --      &        --      &        --      &        --      &  $>$21.92      &     3$\ddag$  \\ 
2018-12-15 &    58467.32  &          --       &        --      &        --      &    $>$20.97    &        --      &       --       &     3  \\ 
2019-08-20 &    58715.57  &          --       &        --      &        --      &    $>$20.57    &        --      &       --       &     3  \\ 
2019-08-24 &    58719.54  &          --       &        --      &        --      &    $>$21.39    &        --      &       --       &     3  \\ 
2019-09-19 &    58745.59  &          --       &        --      &        --      & 22.254 (0.396) &        --      &        --      &     3  \\ 
2019-10-13 &    58769.42  &          --       &        --      &        --      & 21.500 (0.429) &        --      &        --      &     3  \\ 
2019-10-22 &	58778.44  &          --       &        --      &        --      &        --      &        --      &  21.847 (0.108)&     3$\ddag$  \\ 
2019-11-05 & 	58792.43  &          --       &        --      &        --      & 21.176 (0.100) &        --      &        --      &     3  \\ 
2019-11-24 &	58811.15  &          --       &        --      &  $>$20.24      &        --      &        --      &        --      &     4  \\ 
2019-11-24 &	58811.19  &          --       &  $>$19.79      &        --      &        --      &        --      &        --      &     4  \\ 
2019-11-25 &	58812.24  &          --       &        --      &  $>$20.82      &        --      &        --      &        --      &     4  \\ 
2019-11-27 &	58814.11  &          --       &  $>$18.66      &        --      &        --      &        --      &        --      &     4  \\ 
2019-11-27 &	58814.15  &          --       &        --      &  $>$19.93      &        --      &        --      &        --      &     4  \\ 
2019-11-27 &	58814.19  &          --       &        --      &  $>$19.91      &        --      &        --      &        --      &     4  \\ 
2019-11-27 &	58814.30  &          --       &        --      &        --      &        --      &        --      &  21.066 (0.078)&     3$\ddag$  \\ 
2019-11-29 &	58816.32  &          --       &  $>$20.71      &        --      &        --      &        --      &        --      &     5$\dag$   \\ 
2019-11-30 &	58818.41  &          --       &        --      &        --      &        --      &        --      &  $>$18.76      &     6  \\ 
2019-12-02 &	58819.12  &          --       &        --      &  $>$18.89      &        --      &        --      &        --      &     4  \\ 
2019-12-02 &	58819.16  &          --       &        --      &  $>$20.15      &        --      &        --      &        --      &     4  \\ 
2019-12-02 &	58819.19  &          --       &  $>$18.54      &        --      &        --      &        --      &        --      &     4  \\ 
2019-12-03 &	58820.16  &          --       &  $>$19.41      &        --      &        --      &        --      &        --      &     4  \\ 
2019-12-03 &    58820.36  &          --       &        --      &        --      &        --      &        --      &  $>$19.80      &     5  \\ 
2019-12-05 &    58822.29  &          --       &        --      &        --      &        --      &        --      &  20.232 (0.519)&   	 6  \\ 
2019-12-07 &    58824.33  &          --       &        --      &        --      &        --      &        --      &  19.854 (0.259)&	 5  \\ 
2019-12-09 &    58826.34  &          --       &        --      &        --      &        --      &        --      &  20.008 (0.501)&	 6  \\ 
2019-12-10 &    58827.40  &          --       &        --      &        --      &        --      &        --      &  $>$18.81      &     5  \\ 
2019-12-11 &	58828.12  &          --       &  $>$19.10      &        --      &        --      &        --      &        --      &     4  \\ 
2019-12-11 &	58828.19  &          --       &        --      &  $>$18.64      &        --      &        --      &        --      &     4  \\ 
2019-12-11 &	58828.34  &          --       &        --      &        --      &        --      &        --      &  $>$19.41      &     6  \\ 
2019-12-12 &	58829.31  &          --       &        --      &        --      &        --      &        --      &  20.130 (0.474)&	 5  \\ 
2019-12-13 &	58830.18  &          --       &  $>$19.84      &        --      &        --      &        --      &        --      &     4  \\ 
2019-12-13 &    58830.22  &          --       &        --      &        --      &  19.590 (0.117)&        --      &        --      &     3  \\ 
2019-12-13 &	58830.28  &          --       &        --      &  $>$20.02      &        --      &        --      &        --      &     4  \\ 
2019-12-13 &	58830.29  &          --       &        --      &        --      &        --      &        --      &  20.298 (0.451)&     6  \\ 
2019-12-14 &	58831.10  &          --       &        --      &  20.345 (0.243)&        --      &        --      &        --      &     4  \\ 
2019-12-14 &	58831.12  &          --       &  $>$19.41      &        --      &        --      &        --      &        --      &     4  \\ 
2019-12-16 &	58833.36  &          --       &        --      &        --      &        --      &        --      &  20.178 (0.494)&     6  \\ 
2019-12-17 &    58834.19  &          --       &        --      &  19.921 (0.192)&        --      &        --      &        --      &     7  \\ 
2019-12-18 &	58835.14  &          --       &  $>$19.08      &        --      &        --      &        --      &        --      &     4  \\ 
2019-12-19 &    58836.35  &          --       &        --      &        --      &        --      &        --      &  19.964 (0.321)&     5  \\ 
2019-12-20 &	58837.08  &          --       &        --      &  20.116 (0.169)&        --      &        --      &        --      &     4  \\ 
2019-12-20 &	58837.14  &          --       &  $>$20.55      &        --      &        --      &        --      &        --      &     4  \\ 
2019-12-21 &	58838.14  &          --       &  $>$19.52      &        --      &        --      &        --      &        --      &     4  \\ 
2019-12-21 &    58838.27  &          --       &        --      &        --      &        --      &        --      &  20.335 (0.329)&     6  \\ 
2019-12-29 &	58846.12  &          --       &        --      &  19.756 (0.114)&        --      &        --      &        --      &     4  \\ 
2019-12-29 &    58846.26  &          --       &        --      &        --      &        --      &        --      &  19.471 (0.208)&     6  \\ 
2020-01-01 &	58849.08  &          --       &        --      &  19.562 (0.135)&        --      &        --      &        --      &     4  \\ 
2020-01-01 &	58849.09  &          --       &  20.084 (0.138)&        --      &        --      &        --      &        --      &     4  \\ 
2020-01-01 &	58849.09  &          --       &  20.058 (0.178)&        --      &        --      &        --      &        --      &     4  \\ 
2020-01-01 &	58849.15  &          --       &        --      &  19.547 (0.121)&        --      &        --      &        --      &     4  \\ 
2020-01-02 &    58850.30  & 	      --      &        --      &        --      &        --      &        --      &  19.157 (0.192)&     6  \\ 
2020-01-03 &	58851.13  &          --       &  $>$19.13      &        --      &        --      &        --      &        --      &     4  \\ 
2020-01-04 &	58852.11  &          --       &  19.693 (0.315)&        --      &        --      &        --      &        --      &     4  \\ 
2020-01-04 &	58852.16  &          --       &        --      &  19.480 (0.116)&        --      &        --      &        --      &     4  \\ 
2020-01-05 &	58853.08  &          --       &        --      &  $>$19.01      &        --      &        --      &        --      &     4  \\ 
2020-01-06 &	58854.14  &          --       &        --      &  19.312 (0.200)&        --      &        --      &        --      &     4  \\ 
2020-01-06 &    58854.25  & 	      --      &        --      &        --      &        --      &        --      &  18.889 (0.079)&     6  \\ 
2020-01-07 &	58855.16  &          --       &  19.758 (0.236)&        --      &        --      &        --      &        --      &     4  \\ 
2020-01-07 &	58855.19  &          --       &        --      &  19.248 (0.166)&        --      &        --      &        --      &     4  \\ 
2020-01-11 &	58859.10  &          --       &  19.674 (0.254)&        --      &        --      &        --      &        --      &     4  \\ 
2020-01-11 &	58859.18  &          --       &        --      &  19.149 (0.159)&        --      &        --      &        --      &     4  \\ 
2020-01-12 &	58860.18  &          --       &        --      &  18.921 (0.162)&        --      &        --      &        --      &     4  \\ 
2020-01-14 &	58862.11  &          --       &  19.845 (0.124)&        --      &        --      &        --      &        --      &     4  \\ 
2020-01-16 &	58864.11  &          --       &  19.701 (0.129)&        --      &        --      &        --      &        --      &     4  \\ 
2020-01-22 &    58866.31  & 	      --      &        --      &        --      &        --      &        --      &  19.050 (0.087)&     6  \\ 
2020-01-19 &	58867.14  &          --       &  $>$19.25      &        --      &        --      &        --      &        --      &     4  \\ 
2020-01-20 &    58868.27  & 	      --      &  19.665 (0.264)&        --      &        --      &        --      &        --      &     5$\dag$   \\ 
2020-01-22 &    58870.22  & 	      --      &        --      &        --      &        --      &        --      &  19.135 (0.195)&     6  \\ 
2020-01-23 &	58871.09  &          --       &        --      &  19.356 (0.141)&        --      &        --      &        --      &     4  \\ 
2020-01-23 &	58871.16  &          --       &  19.657 (0.198)&        --      &        --      &        --      &        --      &     4  \\ 
2020-01-24 &    58872.23  & 	      --      &  19.636	(0.367)&        --      &        --      &        --      &        --      &     5$\dag$   \\ 
2020-01-26 &	58874.12  &          --       &        --      &  19.314 (0.083)&        --      &        --      &        --      &     4  \\ 
2020-01-26 &    58874.26  & 	      --      &        --      &        --      &        --      &        --      &  19.158 (0.087)&     6  \\ 
2020-01-28 &	58876.26  & 	      --      &        --      &        --      &        --      &        --      &  19.063 (0.194)&     5  \\ 
2020-01-29 &	58877.12  &          --       &        --      &  19.133 (0.128)&        --      &        --      &        --      &     4  \\ 
2020-01-29 &	58877.18  &          --       &  19.502 (0.141)&        --      &        --      &        --      &        --      &     4  \\ 
2020-02-01 &	58880.15  &          --       &        --      &  18.954 (0.097)&        --      &        --      &        --      &     4  \\ 
2020-02-03 &	58882.12  &          --       &  19.601 (0.166)&        --      &        --      &        --      &        --      &     4  \\ 
2020-02-03 &    58882.24  & 	      --      &        --      &        --      &        --      &        --      &  18.736 (0.170)&     6  \\ 
2020-02-05 &	58884.13  &          --       &        --      &  18.813 (0.187)&        --      &        --      &        --      &     4  \\ 
2020-02-05 &    58884.28  & 	      --      &        --      &        --      &        --      &        --      &  18.842 (0.209)&     5  \\ 
2020-02-06 &	58885.41  & 	      --      &        --      &  18.779 (0.400)&        --      &        --      &        --      &     1  \\ 
2020-02-07 &	58886.38  & 	      --      &        --      &  18.755 (0.233)&        --      &        --      &        --      &     1  \\ 
2020-02-08 &	58887.11  &          --       &        --      &  18.717 (0.091)&        --      &        --      &        --      &	 4  \\ 
2020-02-08 &	58887.18  &          --       &  19.183 (0.269)&        --      &        --      &        --      &        --      &	 4  \\ 
2020-02-08 &	58887.41  & 	      --      &        --      &  18.509 (0.416)&        --      &        --      &        --      &     1  \\ 
2020-02-09 &	58888.41  & 	      --      &        --      &  18.210 (0.281)&        --      &        --      &        --      &     1  \\ 
2020-02-11 &	58890.41  & 	      --      &        --      &  16.253 (0.052)&        --      &        --      &        --      &     1  \\ 
2020-02-12 &	58891.11  &          --       &        --      &  15.437 (0.018)&        --      &        --      &        --      &  	 4  \\ 
2020-02-12 &	58891.16  &          --       &  15.389 (0.026)&        --      &        --      &        --      &        --      &	 4  \\ 
2020-02-12 &	58891.39  & 	      --      &        --      &  15.110 (0.028)&        --      &        --      &        --      &     1  \\ 
2020-02-12 &    58892.11  & 	      --      &  15.422	(0.081)&        --      &        --      &        --      &        --      &	 8  \\ 
2020-02-13 &	58892.41  & 	      --      &        --      &  15.062 (0.055)&        --      &        --      &        --      &     1  \\ 
2020-02-13 &	58892.86  & 	      --      &        --      &  15.206 (0.012)&        --      &        --      &        --      & 	 2  \\ 
2020-02-14 &	58893.41  & 	      --      &        --      &  15.654 (0.057)&        --      &        --      &        --      & 	 1  \\ 
2020-02-14 &    58893.09  & 	      --      &  15.836	(0.122)&        --      &        --      &        --      &        --      &	 8  \\ 
2020-02-14 &    58893.14  & 	      --      &        --      & 	--      &  15.621 (0.168)&        --      &        --      &	 9  \\ 
2020-02-14 &	58893.73  &          --       &        --      &  15.670 (0.060)&        --      &        --      &        --      &	 9  \\ 
2020-02-14 &    58893.75  &          --       &	 15.954 (0.012)&  15.731 (0.012)&  15.767 (0.012)&        --      &        --      &     9  \\ 
2020-02-14 &	58893.81  &    16.646 (0.021) &	 15.953 (0.011)&  15.766 (0.016)&  15.817 (0.017)&  15.949 (0.022)&        --      &   	10  \\ 
2020-02-15 &	58894.13  &          --       &        --      &  16.128 (0.032)&        --      &        --      &        --      &	 4  \\ 
2020-02-15 &	58894.16  &          --       &  16.247 (0.034)&        --      &        --      &        --      &        --      &	 4  \\ 
2020-02-15 &	58894.74  &          --       &  16.568 (0.037)&  16.349 (0.059)&  16.289 (0.085)&        --      &        --      &    11  \\ 
2020-02-15 &	58894.79  &    17.680 (0.025) &	 16.578 (0.011)&  16.343 (0.015)&  16.291 (0.016)&  16.312 (0.014)&        --      &	10  \\ 
2020-02-15 &	58894.88  &    17.726 (0.039) &        --      &  16.368 (0.010)&  16.294 (0.009)&        --      &        --      &	2   \\ 
2020-02-16 &	58895.86  &    18.502 (0.049) &        --      &  16.496 (0.012)&  16.371 (0.010)&        --      &        --      &	2   \\ 
2020-02-17 &	58896.83  &    18.724 (0.099) &  17.129 (0.015)&  16.522 (0.011)&  16.381 (0.010)&  16.361 (0.014)&        --      &  	2   \\ 
2020-02-17 &	58896.84  &    18.767 (0.045) &  17.136 (0.009)&  16.527 (0.009)&  16.389 (0.008)&  16.364 (0.014)&        --      &	12  \\ 
2020-02-18 &	58897.42  & 	      --      &        --      &  16.538 (0.063)&        --      &        --      &        --      &     1  \\ 
2020-02-18 &	58897.83  & 	      --      &        --      &  16.544 (0.020)&        --      &        --      &        --      & 	10$\star$  \\ 
2020-02-19 &	58898.42  & 	      --      &        --      &  16.563 (0.063)&        --      &        --      &        --      & 	 1  \\ 
2020-02-20 &	58899.42  & 	      --      &        --      &  16.561 (0.067)&        --      &        --      &        --      & 	 1  \\ 
2020-02-21 &    58900.16  &          --       &        --      &  16.571 (0.047)&        --      &        --      &        --      &	 4  \\ 
2020-02-21 &    58900.23  &          --       &  17.262	(0.096)&        --      &        --      &        --      &        --      & 	 5$\dag$   \\ 
2020-02-21 &	58900.42  & 	      --      &        --      &  16.566 (0.081)&        --      &        --      &        --      & 	 1  \\ 
2020-02-21 &	58900.85  & 	      --      &        --      &  16.568 (0.021)&        --      &        --      &        --      & 	12$\star$  \\ 
2020-02-22 &	58901.42  & 	      --      &        --      &  16.579 (0.080)&        --      &        --      &        --      & 	 1  \\ 
2020-02-23 &	58902.43  & 	      --      &        --      &  16.601 (0.067)&        --      &        --      &        --      &     1  \\ 
2020-02-24 &	58903.42  & 	      --      &        --      &  16.646 (0.069)&        --      &        --      &        --      &     1  \\ 
2020-02-25 &    58904.83  &    20.667 (0.467) &        --      &        --      &        --      &        --      &        --      & 	12  \\ 
2020-02-26 &    58905.12  & 	      --      &        --      &  16.538 (0.038)&        --      &        --      &        --      &	 4  \\ 
2020-02-26 &    58905.14  &  	      --      &  17.865 (0.095 &        --      &        --      &        --      &        --      &	 4  \\ 
2020-02-26 &	58905.46  & 	      --      &        --      &  16.738 (0.265)&        --      &        --      &        --      &     1  \\ 
2020-02-27 &	58906.79  &  	      --      &        --      &  16.686 (0.019)&  16.282 (0.019)&  16.140 (0.026)&        --      &	10  \\ 
2020-02-28 &	58907.74  & 	      --      &  18.259 (0.050)&  16.786 (0.021)&  16.306 (0.019)&  16.180 (0.027)&        --      &	10  \\ 
2020-03-02 &	58910.43  & 	      --      &        --      &  16.977 (0.105)&        --      &        --      &        --      &     1  \\ 
2020-03-03 &	58911.83  & 	      --      &  18.818 (0.117)&  17.030 (0.033)&  16.441 (0.022)&  16.202 (0.019)&        --      &	2   \\ 
2020-03-05 &	58913.84  & 	      --      &        --      &  17.174 (0.066)&        --      &        --      &        --      &    12$\star$  \\ 
2020-03-06 &	58914.44  & 	      --      &        --      &  17.265 (0.211)&        --      &        --      &        --      & 	 1  \\ 
2020-03-07 &	58915.78  & 	      --      &  19.283 (0.205)&  17.384 (0.072)&  16.629 (0.038)&        --      &        --      & 	11  \\ 
2020-03-07 &	58915.83  & 	      --      &        --      &        --      &        --      &  16.277 (0.015)&        --      &	2   \\ 
2020-03-08 &	58916.45  &          --       &        --      &  17.527 (0.242)&        --      &        --      &        --      &     1  \\ 
2020-03-08 &	58916.76  & 	      --      &  19.408 (0.283)&  17.543 (0.036)&  16.729 (0.048)&        --      &        --      & 	11  \\ 
2020-03-11 &	58919.78  & 	      --      &  19.972 (0.373)&  17.985 (0.061)&  16.944 (0.033)&        --      &        --      &    11  \\ 
2020-03-12 &    58920.85  &          --       &        --      &  18.198 (0.185)&        --      &        --      &        --      &    12$\star$  \\ 
2020-03-16 &	58924.76  & 	      --      &        --      &        --      &  17.570 (0.273)&        --      &        --      &    11  \\ 
2020-03-17 &	58925.84  &          --       &        --      &  19.046 (0.100)&        --      &  17.087 (0.050)&        --      &    13  \\ 
2020-03-18 &	58926.77  & 	      --      &        --      &  19.329 (0.402)&  17.906 (0.570)&        --      &        --      &    11  \\ 
2020-03-18 &	58926.77  & 	      --      &  $>$21.14      &        --      &        --      &        --      &        --      &    11  \\ 
\hline                                                                                                                   
\end{longtable}                                                                                                          
                                                                                                                         
\tablefoot{                                                                                                              
1 = 35 cm F/11 telescope + KAF-1001E CCD (Itagaki Astronomical Observatory, Yamagata, Japan);                     
2 = 2.0 m Liverpool Telescope (LT) + IO:O (La Palma, Canary Islands, Spain);                                            
3 = 1.8 m Pan-STARRS Telescopes + GPC cameras (Haleakala, Hawaii Islands, USA);        
4 = 1.2 m S. Oschin Telescope + ZTF-Cam (Mt. Palomar, USA);
5 = 0.5 m ATLAS Telescope + ACAM2 (Mauna Loa, Hawaii Islands, USA);                                            
6 = 0.5 m ATLAS Telescope + ACAM1 (Haleakala, Hawaii Islands, USA);                                          
7 = 2.0m Faulkees North Telescope + fa05 camera (Hawaii Isl., USA);
8 = ASAS-SN 4$\times$0.16 m Brutus Telescope + FLI ProLine PL230 CCD (LCOGT - Haleakala, Hawaii Islands, USA; see \protect\citet{koc17});          
9 = $B,V,R,I$ Johnson-Bessell-Cousins observations from AAVSO, TOCP and ANS collaboration (see, also, Table \ref{tabA1}), converted in Sloan bands following   the prescriptions of \protect\citet{jor06};
10 = 1.82 m Copernico Telescope + AFOSC (Cima Ekar, Asiago, Italy);                                                   
11 = 67/92 cm Schmidt Telescope + Moravian G4-16000LC + KAF-16803 CCD CCD (Cima Ekar, Asiago, Italy);    
12 = 2.56 m Nordic Optical Telescope (NOT) + ALFOSC (La Palma, Canary Islands, Spain);                                   
13 = 2.56 m Nordic Optical Telescope (NOT) + StanCam (La Palma, Canary Islands, Spain).

Additional information: 
$\ddag$ Pan-STARRS $w$-band data;
$\dag$ ATLAS-$c$ band data, converted to Sloan-$g$ \protect\citep{ton18};
$\star$ Unfiltered pointing data, scaled to Sloan-$r$ photometry.}                                                                                                                \end{landscape}                                                                                                          
}

\end{appendix}

\end{document}